\documentclass[11pt,letterpaper]{amsart}
\usepackage{graphicx}
\usepackage{xspace}
\usepackage{hyperref}
\usepackage{amsmath}
\usepackage{amssymb}
\usepackage{amsthm}
\usepackage{euscript}
\usepackage[top=1in,bottom=1in,left=1in,right=1in]{geometry}

\theoremstyle{plain}
\newtheorem{theorem}{Theorem}
\newtheorem{lemma}[theorem]{Lemma}
\newtheorem{proposition}[theorem]{Proposition}
\newtheorem{corollary}[theorem]{Corollary}
\theoremstyle{definition}

\newtheorem{example}{Example}
\theoremstyle{remark}
\newtheorem*{remark}{Remark}

\newcommand{\veps}{\ensuremath{\varepsilon}\xspace}
\newcommand{\uclos}{\ensuremath{\mathord{\uparrow}}\xspace}
\newcommand{\inv}{^{-1}}
\newcommand{\nat}{\ensuremath{\mathbb{N}}\xspace}
\newcommand{\frB}{\ensuremath{\mathbb{B}}\xspace}
\newcommand{\frP}{\ensuremath{\mathbb{P}}\xspace}

\newcommand{\Bs}{\ensuremath{\mathcal{B}}\xspace}
\newcommand{\Cs}{\ensuremath{\mathcal{C}}\xspace}
\newcommand{\Ds}{\ensuremath{\mathcal{D}}\xspace}

\newcommand{\Gs}{\ensuremath{\mathcal{G}}\xspace}

\newcommand{\Hb}{\ensuremath{\mathbf{H}}\xspace}

\newcommand{\Kb}{\ensuremath{\mathbf{K}}\xspace}

\newcommand{\Pb}{\ensuremath{\mathbf{P}}\xspace}

\newcommand{\Ub}{\ensuremath{\mathbf{U}}\xspace}

\newcommand{\wsuit}{well-suited\xspace}

\newcommand{\vari}{prevariety\xspace}

\newcommand{\varis}{prevarieties\xspace}

\newcommand{\pvari}{positive prevariety\xspace}

\newcommand{\bool}[1]{\ensuremath{Bool(#1)}\xspace}
\newcommand{\pol}[1]{\ensuremath{Pol(#1)}\xspace}
\newcommand{\bpol}[1]{\ensuremath{BPol(#1)}\xspace}

\newcommand{\booln}{\ensuremath{Bool}\xspace}
\newcommand{\poln}{\ensuremath{Pol}\xspace}
\newcommand{\bpoln}{\ensuremath{BPol}\xspace}

\newcommand{\grp}{\ensuremath{\textup{GR}}\xspace}
\newcommand{\abg}{\ensuremath{\textup{AMT}}\xspace}
\newcommand{\md}{\ensuremath{\textup{MOD}}\xspace}
\newcommand{\stzer}{\textup{ST}\xspace}

\newcommand{\bsc}[1]{\ensuremath{\Bs\Sigma_{#1}}\xspace}
\newcommand{\bsg}{\ensuremath{\bsc{1}(<,\frP_{\Gs})}\xspace}
\newcommand{\bsgp}{\ensuremath{\bsc{1}(<,+1,\frP_{\Gs})}\xspace} 
\begin{document}

\title{Characterizing level one in group-based concatenation hierarchies}
\author{Thomas Place \and Marc Zeitoun}
\thanks{Funded by the DeLTA project (ANR-16-CE40-0007).}
\email{firstname.name@labri.fr}
\address{Univ. Bordeaux, CNRS, Bordeaux INP, LaBRI, UMR 5800, F-33400 Talence, France}

\begin{abstract}
  We investigate two operators on classes of regular languages: polynomial closure (\poln) and Boolean closure (\booln). We apply these operators to classes of \emph{group languages} \Gs and to their \wsuit extensions $\Gs^+$, which is the least Boolean algebra containing $\Gs$ and $\{\veps\}$. This yields the classes \bool{\pol{\Gs}} and \bool{\pol{\Gs^+}}. These classes form the first level in important classifications of classes of regular languages, called \emph{concatenation hierarchies}, which admit natural \emph{logical characterizations}. We present \emph{generic} algebraic characterizations of these classes. They imply that one may decide whether a regular language belongs to such a class, provided that a more general problem called separation is decidable for the input class \Gs. The proofs are constructive and rely exclusively on notions from language and automata theory.
\end{abstract}

\maketitle

\section{Introduction}\label{sec:intro}

An active line of research in automata theory is to investigate natural subclasses of the regular languages of finite words. We are particularly interested in classes associated to fragments of standard pieces of syntax used to define the regular languages (\emph{e.g.}, regular expressions or monadic second-order logic). Given a fragment, we consider the class of all languages that can be defined by an expression of this fragment. For each such class \Cs, a standard approach for its investigation is to look for a \emph{\Cs-membership algorithm}: given a regular language $L$ as input, decide whether $L\in\Cs$. Getting such an algorithm requires a solid understanding of~\Cs. We are not only interested in~a yes/no answer on the decidability of \Cs-membership but also in the techniques and proof arguments involved in order to obtain this answer.

We look at classifications called \emph{concatenation hierarchies}. A concatenation hierarchy is built from an input class of languages, called its \emph{basis}, using two operators. The \emph{polynomial closure} of a class \Cs, written \pol{\Cs}, consists in all finite unions of languages $L_0a_1L_1 \cdots a_nL_n$ where $a_1,\dots,a_n$ are letters and $L_0,\dots,L_n$ are languages in~\Cs. The \emph{Boolean closure} of \Cs, denoted by \bool{\Cs}, is the least class containing \Cs and closed under Boolean operations. We investigate level \emph{one} of concatenation hierarchies: the classes \bool{\pol{\Cs}} (abbreviated \bpol{\Cs}). Moreover, we consider special bases \Cs. The \emph{group languages} are those  recognized by a finite group, or equivalently by a permutation automaton (\emph{i.e.}, a complete, deterministic \emph{and} co-deterministic automaton). We only consider bases that are either a class \Gs containing only group languages, or its \wsuit extension $\Gs^+$ (roughly, $\Gs^+$ is the least Boolean algebra containing~\Gs and the singleton $\{\veps\}$). The motivation for using such bases stems from the logical characterizations of concatenation hierarchies~\cite{ThomEqu,PZ:generic18}. A word can be viewed as a logical structure consisting of a sequence of \emph{labeled positions}. Therefore, we may use first-order sentences to define languages. It turns out that \bpol{\Gs} and \bpol{\Gs^+} correspond to the logical classes \bsg and \bsgp where \bsc{1} is the fragment of first-order logic containing only the Boolean combinations of purely existential formulas. Here, the predicates ``$<$'' and ``$+1$'' are interpreted as the linear order and the successor relation. Moreover, $\frP_{\Gs}$ is a set of predicates built from \Gs: for each language $L\in\Gs$, it contains a unary predicate that checks whether the prefix preceding a given position belongs to $L$.

In the paper, we present generic algebraic characterizations of \bpol{\Gs} and \bpol{\Gs^+}. They apply to all classes of group languages \Gs satisfying mild hypotheses (namely, \Gs must be closed under Boolean operations and quotients). Moreover, they imply that \emph{membership} is decidable for both \bpol{\Gs} and \bpol{\Gs^+} provided that a more general problem, \emph{separation}, is decidable for \Gs. Separation takes two input regular languages $L_0,L_1$ and asks whether there exists $K \in \Gs$ such that $L_0 \subseteq K$ and $L_1 \cap K = \emptyset$. From the decidability point of view, the results are not entirely new. In particular, for \bpol{\Gs}, it is even known~\cite{pzconcagroup} that \emph{separation} is decidable for \bpol{\Gs} when it is already decidable for \Gs (on the other hand, this is open for \bpol{\Gs^+}). Hence, our main contribution consists in the characterizations themselves and the techniques that we use to prove them. In particular, the proof arguments are constructive. For example, given a language $L$ satisfying the characterization of  \bpol{\Gs}, we prove directly that $L$ belongs to \bpol{\Gs} by explicitly building a description of $L$ as a Boolean combination of products $L_0a_1L_1 \cdots a_nL_n$ where $L_0,\dots,L_n \in \Gs$.

With these characterizations, we generalize a number of known results for particular classes of group languages \Gs. Let us first consider the case when \Gs is the trivial Boolean algebra, which we denote by \stzer: we have $\stzer = \{\emptyset,A^*\}$ and $\stzer^+ = \{\emptyset,\{\veps\},A^+,A^*\}$ (where $A$ is the alphabet). In this case, we obtain two well-known classes: $\bpol{\stzer}= \bsc{1}(<)$ defines the \emph{piecewise testable languages} and $\bpol{\stzer^+}=\bsc{1}(<,+1)$ the \emph{languages of dot-depth one}. The famous algebraic characterizations of these classes by Simon~\cite{simonthm} and Knast~\cite{knast83} are simple corollaries of our generic results. Another key example is the class \md of \emph{modulo languages}: membership of a word in such a language depends only on its length modulo some fixed integer. In this case, the logical counterparts of \bpol{\md} and \bpol{\md^+} are the classes $\bsc{1}(<,MOD)$ and $\bsc{1}(<,+1,MOD)$ where ``$\mathit{MOD}$'' denotes the set of \emph{modular predicates}. It is again possible to use our results to reprove the known characterizations of these classes by Chaubard, Pin and Straubing~\cite{ChaubardPS06} and Maciel, Péladeau and Thérien~\cite{MACIEL2000135}. Our result also applies to the important case when \Gs is the class \grp of \emph{all} group languages~\cite{margpin85}. In particular, there exists a specialized characterization of \bpol{\grp} by Henckell, Margolis, Pin and Rhodes~\cite{henckell:hal-00019815} which is independent from \grp-separation. While it is also possible to reprove this result as a corollary of our characterization, this requires a bit of technical work as well as knowledge of the \grp-separation algorithm~\cite{Ash91} which is a difficult result. Finally, another generic characterization of the classes \bpol{\Gs} follows from an algebraic theorem of Steinberg~\cite{Steinberg01} (though it only applies under more restrictive hypotheses on \Gs).

The techniques used in the paper are quite different from those used for proving the aforementioned specialized results. Historically, classes of the form \bpol{\Gs} or \bpol{\Gs^+} are often approached via alternate definitions based on an algebraic construction called ``\emph{wreath product}''. Indeed, it turns out that all classes of this kind can be built from the piecewise testable languages (\emph{i.e.}, the class \bpol{\stzer}) using this product~\cite{StrauVD,pin:hal-00112635}. The arguments developed in~\cite{ChaubardPS06,MACIEL2000135,margpin85,henckell:hal-00019815,Steinberg01} build exclusively on this construction. The paper is completely independent from these techniques: we work \emph{directly} with the language theoretic definition of our classes based on the operator \bpoln. This matches our original motivation: investigating classes of regular languages.

\smallskip

We introduce the terminology that we shall need in Section~\ref{sec:prelims}. Then, we look at classes of the form \bpol{\Gs} in Section~\ref{sec:bpolg}. Finally, we devote Section~\ref{sec:bpolgp} to classes of the form \bpol{\Gs^+}.

\section{Preliminaries}
\label{sec:prelims}
In this section, we present the objects that we investigate in the paper and introduce the terminology that we require in order to manipulate them.

\subsection{Words, regular languages and classes}

We fix an arbitrary finite alphabet $A$ for the whole paper. As usual, $A^*$ denotes the set of all finite words over $A$, including the empty word \veps. We let $A^+ = A^* \setminus \{\veps\}$. For $u,v \in A^*$, we let $uv$ be the word obtained by concatenating $u$ and $v$. Additionally, given $w\in A^*$, we write $|w|\in\nat$ for the length of~$w$. A language is a subset of $A^*$. We denote the singleton language $\{u\}$ by $u$. We lift concatenation to languages: for $K,L \subseteq A^*$, we let $KL = \{uv \mid u \in K \text{ and } v \in L\}$. We shall consider \emph{marked products}: given languages $L_0,\dots,L_n \subseteq A^*$, a marked product of $L_0,\dots,L_n$ is a product of the form $L_0a_1L_1 \cdots a_nL_n$ where $a_1,\dots,a_n \in A$ (note that ``$L_0$'' is a marked product: this is the case $n =0$).

\smallskip
\noindent
{\bf Regular languages.} All languages considered in the paper are \emph{regular}. These are the languages that can be equivalently defined by a regular expression, an automaton or a morphism into a finite monoid. We work with the latter definition. A \emph{monoid} is a set $M$ equipped with a binary operation \mbox{$s,t \mapsto st$} (called multiplication) which is associative and has a neutral element denoted by ``$1_M$''. Recall that an idempotent of a monoid $M$ is an element $e \in M$ such that $ee = e$. For all $S \subseteq M$, we write $E(S)$ for the set of all idempotents in $S$. It is standard that when $M$ is \emph{finite}, there exists $\omega(M) \in \nat$ (written $\omega$ when $M$ is understood) such that $s^\omega$ is idempotent for every $s \in M$.

An \emph{ordered monoid} is a pair $(M,\leq)$ where $M$ is a monoid and $\leq$ is a partial order on $M$ which is compatible with multiplication: for every $s,t,s',t' \in M$, if $s \leq t$ and $s' \leq t'$, then $ss' \leq tt'$. A upper set of $M$ (for $\leq$) is a set $S \subseteq M$ which is upward closed for $\leq$: for every $s,t \in M$ such that $s \leq t$, we have $s \in S \Rightarrow t \in S$. For every $s \in M$, we write $\uclos s$ for the least upper set of $M$ containing~$s$ (\emph{i.e.}, $\uclos s$ consists of all $t \in M$ such that $s \leq t$). We may view arbitrary monoids as being ordered, as follows: we view any monoid $M$ with no ordering specified as the ordered monoid $(M,=)$: we use equality as the ordering. In this special case, \emph{all} subsets of $M$ are upper sets.

Clearly, $A^*$ is a monoid for concatenation as the multiplication (\veps is neutral).  Given an ordered monoid $(M,\leq)$, we may consider morphisms $\alpha: A^* \to (M,\leq)$. We say that a language $L \subseteq A^*$ is \emph{recognized} by such a morphism $\alpha$ when there exists a \emph{upper set} $F\subseteq M$ such that $L=\alpha\inv(F)$ (the definition depends on the ordering $\leq$, since $F$ must be a upper set). Note that this also defines the languages recognized by a morphism $\eta: A^* \to N$ into a \emph{unordered} monoid $N$ since we view $N$ as the ordered monoid $(N,=)$. It is well-known that a language is regular if and only if it can be recognized by a morphism into a \emph{finite} monoid.

\begin{remark} \label{rem:finmono}
  The only infinite monoid that we consider is $A^*$. From now, we implicitly assume that every other monoid $M,N,\dots$ that we consider is finite.
\end{remark}

\noindent
{\bf Classes of languages.} A class of languages \Cs is a set of languages. A \emph{lattice} is a class closed under both union and intersection, and containing the languages $\emptyset$ and $A^*$. Moreover, a \emph{Boolean algebra} is a lattice closed under complement. Finally, a class \Cs is \emph{quotient-closed} when for all $L \in \Cs$ and $u,v \in A^*$, the language $\{w \in A^* \mid uwv \in L\}$ belongs to \Cs as well. We say that a class \Cs is a \emph{\pvari} (resp. \emph{\vari}) to indicate that it is a quotient-closed lattice (resp. Boolean algebra) containing \emph{only regular languages}.

We rely on a decision problem called \emph{membership} as a means to investigate classes of languages. Given a class \Cs, the \Cs-membership problem takes as~input a regular language $L$ and asks whether $L \in \Cs$. Intuitively, obtaining a procedure for \Cs-membership requires a solid understanding of \Cs. We also look at more involved problem called \emph{separation}. Given a class \Cs, and two languages $L_0$ and $L_1$, we say that $L_0$ is \Cs-separable from $L_1$ if and only if there exists $K \in \Cs$ such that $L_0 \subseteq K$ and $L_1 \cap K = \emptyset$. The \Cs-separation problem takes two regular languages $L_0$ and $L_1$ as input and asks whether $L_0$ is \Cs-separable from $L_1$. Let us point out that we do \emph{not} present separation algorithms in this paper. We shall need this problem as an intermediary in our investigation of membership.

\smallskip
\noindent
{\bf Group languages.} A group is a monoid $G$ such that every element $g \in G$ has an inverse $g\inv \in G$, \emph{i.e.}, such that \hbox{$gg\inv = g\inv g = 1_G$}. We call ``\emph{group language}'' a language recognized by a morphism into a \emph{finite group}. We consider classes \Gs that are \varis of group languages (\emph{i.e.}, containing group languages only).

We let \grp as the class of \emph{all} group languages. Another important example is the class \abg of \emph{alphabet modulo testable languages}. For every $w\in A^*$ and every $a \in A$, we write $\#_a(w) \in \nat$ for the number of occurrences of ``$a$'' in $w$. The class \abg consists in all finite Boolean combinations of languages $\{w \in A^* \mid \#_a(w) \equiv k \bmod m\}$ where $a \in A$ and $k,m \in \nat$ such that $k < m$. One may verify that these are exactly the languages recognized by commutative groups. We also consider the class \md, which consists in all finite Boolean combinations of languages $\{w \in A^* \mid |w| \equiv k \bmod m\}$ with $k,m \in \nat$ such that $k < m$. Finally, we write \stzer for the trivial class $\stzer = \{\emptyset,A^*\}$. One may verify that \grp, \abg, \md and \stzer are all \varis of group languages.

It follows from the definition that $\{\veps\}$ and $A^+$ are \emph{not} group languages. This motivates the next definition: for a class \Cs, the \emph{\wsuit extension of \Cs}, denoted by $\Cs^+$, consists of all languages of the form $L \cap A^+$ or $L \cup \{\veps\}$ where $L \in \Cs$. The following lemma can follows from the definition.

\begin{lemma} \label{lem:wsuit}
  Let \Cs be a \vari. Then, $\Cs^+$ is a \vari containing the languages $\{\veps\}$ and $A^+$.
\end{lemma}

\subsection{Polynomial and Boolean closure}

In the paper, we look at classes built using two standard operators. Consider a class~\Cs. The \emph{Boolean closure} of \Cs, denoted by \bool{\Cs} is the least Boolean algebra that contains \Cs. Moreover, the \emph{polynomial closure} of \Cs, denoted by \pol{\Cs}, contains all finite unions of marked products $L_0a_1L_1 \cdots a_nL_n$ where $L_0,\dots,L_n  \in \Cs$. Finally, we write \bpol{\Cs} for \bool{\pol{\Cs}}. It is known that when \Cs is a \vari, \pol{\Cs} is a \pvari and \bpol{\Cs} is a \vari. This is not immediate (proving that \pol{\Cs} is closed under intersection is difficult). This was first shown by Arfi~\cite{arfi87}, see also~\cite{jep-intersectPOL,PZ:generic18} for more recent proofs.

\begin{theorem} \label{thm:bpolvar}
  If \Cs is a \vari, then \pol{\Cs} is a \pvari and \bpol{\Cs} is a \vari.
\end{theorem}

In the literature, these operators are used to define classifications called concatenation hierarchies. Given a \vari \Cs, the concatenation hierarchy of basis \Cs is built from \Cs by iteratively applying \poln and \booln to \Cs. In the paper, we only look at the classes \pol{\Cs} and \bpol{\Cs}. These are the levels 1/2 and one in the concatenation hierarchy of basis \Cs. Moreover, we look at bases that are either a \vari of group languages \Gs or its \wsuit extension $\Gs^+$. Most of the prominent concatenation hierarchies in the literature use bases of this kind.

The hierarchy of basis $\stzer = \{\emptyset,A^*\}$ is called the Straubing-Thérien hierarchy~\cite{StrauConcat,TheConcat}. In particular, \bpol{\stzer} is the class of piecewise testable languages~\cite{simonthm}. Another prominent example is the basis $\stzer^+ = \{\emptyset,\{\veps\},A^+,A^*\}$ which yields the \emph{dot-depth hierarchy}~\cite{BrzoDot}. Non-trivial \varis of group languages also yield important hierarchies. For example, the group hierarchy, whose basis is \grp was first investigated in~\cite{margpin85}. The hierarchies of bases \md and $\md^+$ are also prominent (see for example~\cite{ChaubardPS06,MACIEL2000135}). These hierarchies are also interesting for their logical counterparts, which were first discovered by Thomas~\cite{ThomEqu}. Let us briefly recall them (see~\cite{PZ:generic18,pzconcagroup} for more details).

Consider a word $w = a_1 \cdots a_{|w|} \in A^*$. We view $w$ as a linearly ordered set of $|w|+2$ positions $\{0,1,\dots,|w|,|w|+1\}$ such that each position $1 \leq i \leq |w|$ carries the label $a_i \in A$ (on the other hand, $0$ and $|w|+1$ are artificial unlabeled leftmost and rightmost positions). We use first-order logic to describe properties of words: a sentence can quantify over the positions of a word and use a predetermined set of predicates to test properties of these positions. We also allow two constants ``$min$'' and ``$max$'', which we interpret as the artificial unlabeled positions $0$ and $|w|+1$ in a given word $w$. Each first-order sentence $\varphi$ defines the language of all words satisfying the property stated by $\varphi$. Let us present the predicates that we use. For each $a \in A$, we associate a unary predicate (also denoted by~$a$), which selects the positions labeled by ``$a$''. We also consider two binary predicates: the (strict) linear order ``$<$'' and the successor relation~``$+1$''.

\begin{example}
  The sentence ``$\exists x \exists y\ (x < y) \wedge a(x) \wedge b(y)$'' defines the language $A^*aA^*bA^*$. The sentence ``$\exists x \exists y\ a(x) \wedge c(y) \wedge (y+1 = max)$'' defines $A^*aA^*c$.
\end{example}

We associate a (possibly infinite) set of predicates $\frP_{\Gs}$ to every \vari of group languages~\Gs. For every language $L \in \Gs$, $\frP_{\Gs}$ contains a unary predicate $P_L$ which is interpreted as follows. Let $w = a_1 \cdots a_{|w|} \in A^*$. The unary predicate $P_L$ selects all positions $i \in \{0,\dots,|w|+1\}$ such that $i \neq 0$ and $a_1 \cdots a_{i-1} \in L$. It is standard to write ``\bsc{1}'' for the fragment of first-order logic, containing exactly the Boolean combinations of existential first-order sentences. We let \bsg be the class of all languages  defined by a sentence of \bsc{1} using only the label predicates, the linear order ``$<$'' and those in $\frP_{\Gs}$. Moreover, we let \bsgp be the class of all languages defined by a sentence of \bsc{1}, which additionally allows the successor predicate ``$+1$''. The following proposition follows from the generic logical characterization of concatenation hierarchies presented in~\cite{PZ:generic18} and the properties of group languages.

\begin{proposition} \label{prop:logcar}
  For every \vari of group languages \Gs, we have $\bpol{\Gs} = \bsg$ and $\bpol{\Gs^+} = \bsgp$.
\end{proposition}

\begin{remark}
  When $\Gs=\stzer$, all predicates in $\frP_{\stzer}$ are trivial. Hence, we get the classes $\bsc{1}(<)$ and $\bsc{1}(<,+1)$. When $\Gs = \md$, one may verify that we obtain the classes $\bsc{1}(<,MOD)$ and $\bsc{1}(<,+1,MOD)$ where ``$MOD$'' is the set of \emph{modular predicates} (for all $r,q \in \nat$ such that $r < q$, it contains a unary predicate $M_{r,q}$ selecting the positions $i$ such that $i \equiv r \bmod q$). When $\Gs = \abg$, one may verify that we obtain the classes $\bsc{1}(<,AMOD)$ and $\bsc{1}(<,+1,AMOD)$ where ``$AMOD$'' is the set of \emph{alphabetic modular predicates} (for all $a \in A$ and $r,q \in \nat$ such that $r < q$, it contains a unary predicate $M^a_{r,q}$ selecting the positions $i$ such the that number of positions $j < i$ with label $a$ is congruent to $r$ modulo $q$).
\end{remark}

We complete the presentation with a key ingredient of our proofs~\cite[Lemma~3.6]{pzbpolcj}. It describes a concatenation principle for classes of the form \bpol{\Cs}. It is based on the notion of ``cover''. Given a language $L$, a cover of $L$ is a \emph{finite} set \Kb of languages satisfying $L \subseteq \bigcup_{K \in \Kb} K$. If \Ds is a class, we say that \Kb is a \Ds-cover of $L$, if \Kb is a cover of $L$ such that $K \in \Ds$ for every $K \in \Kb$.

\begin{proposition} \label{prop:bconcat}
  Let \Cs be a \vari, and let $n \in \nat$, $L_0,\dots,L_n \in \pol{\Cs}$ and $a_1,\dots,a_n \in A$. For every $i \leq n$, let $\Hb_i$ be a \bpol{\Cs}-cover of $L_i$. There exists a \bpol{\Cs}-cover \Kb of $L_0a_1L_1 \cdots a_nL_n$ such that for every $K \in \Kb$, there exists $H_i \in \Hb_i$ for each $i \leq n$ satisfying $K \subseteq H_0a_1H_1 \cdots a_nH_n$.
\end{proposition}

Moreover, we have the following simple corollary. It will be useful when dealing with classes of the form \bpol{\Cs^+}.

\begin{corollary} \label{cor:bconcat}
  Let \Cs be a \vari, let $L \in \pol{\Cs^+}$, \Hb be a \bpol{\Cs^+}-cover of $L$, $n \in \nat$ and $n+1$ nonempty words \mbox{$w_1,\dots,w_{n+1} \in A^+$}. There exists a \bpol{\Cs^+}-cover \Kb of $w_1L \cdots w_nLw_{n+1}$ such that for each $K \in \Kb$, we have $K \subseteq w_1H_1 \cdots w_nH_nw_{n+1}$ for $H_1,\dots,H_n \in \Hb$.
\end{corollary}

\begin{proof}
  By definition, we have $\{\veps\} \in \Cs^+$. Hence, we may view $w_1L \cdots w_nLw_{n+1}$ as a language of the form $L_0a_1L_1 \cdots a_nL_n$ where each language $L_i$ is either $\{\veps\} \in \pol{\Cs^+}$ or $L \in \pol{\Cs^+}$. Therefore, since $\Hb_\veps = \{\{\veps\}\}$ is a \bpol{\Cs^+}-cover of $\{\veps\}$, we may apply Proposition~\ref{prop:bconcat} to get the desired  \bpol{\Cs^+}-cover \Kb of $w_1L \cdots w_nLw_{n+1}$.
\end{proof}

\subsection{\Cs-morphisms}

We now introduce a key tool, which we shall use to formulate our results. Let \Cs be a \pvari. A \emph{\Cs-morphism} is a \emph{surjective} morphism $\eta: A^*\to (N,\leq)$ into a finite ordered monoid such that every language recognized by $\eta$ belongs to \Cs. Let us make a key remark: when \Cs is a \vari, it suffices to consider \emph{unordered} monoids (we view them as monoids ordered by equality).

\begin{lemma} \label{lem:cmorphbool}
  Let \Cs be a \vari and $\eta: A^* \to (N,\leq)$ a morphism. Then, $\eta$ is a \Cs-morphism if and only if  $\eta: A^* \to (N,=)$ is a \Cs-morphism.
\end{lemma}

\begin{proof}
  The ``if'' direction is trivial. We focus on the converse one. Assume that $\eta$ is a \Cs-morphism. We show that $\eta\inv(s) \in \Cs$ for every $s\in M$. Clearly, this implies that $\eta: A^* \to (N,=)$ is a \Cs-morphism by closure under union. Let $T\subseteq M$ be the set of all elements $t\in M$ such that $s\leq t$ and $s \neq t$. One may verify that we have $\{s\} = (\uclos s) \setminus \left(\bigcup_{t \in T} \uclos t\right)$. Consequently, we obtain $\eta\inv(s) = \eta\inv(\uclos s) \setminus  \left(\bigcup_{t \in T} \eta\inv(\uclos t)\right)$. By hypothesis, this implies that $\eta\inv(s)$ is a Boolean combination of languages in \Cs. We conclude that $\eta\inv(s) \in \Cs$ since \Cs is a Boolean algebra. \qed
\end{proof}

While simple, this notion is a key tool in the paper. First, it is involved in the membership problem. It is well-known that for every regular language $L$, there exists a canonical morphism $\alpha_L: A^* \to (M_L,\leq_L)$ into a finite ordered monoid recognizing $L$ and called the \emph{syntactic morphism} of $L$ (we do not recall the definition as we shall not use it, see~\cite{pingoodref} for example). It can be computed from any representation of $L$ and we have the following standard property.

\begin{proposition} \label{prop:synmemb}
  Let \Cs be a \pvari. A regular language~$L$ belongs to~\Cs if and only if its syntactic morphism $\alpha_L: A^* \to (M_L,\leq_L)$ is a \Cs-morphism.
\end{proposition}

In view of Proposition~\ref{prop:synmemb}, getting an algorithm for \Cs-membership boils down to finding a procedure to decide whether an input morphism $\alpha: A^* \to (M,\leq)$ is a \Cs-morphism. This is how we approach the question in the paper. We shall also use \Cs-morphisms as mathematical tools in proof arguments. In this context, we shall need the following simple corollary of Proposition~\ref{prop:synmemb}.

\begin{proposition}\label{prop:genocm}
  Let \Cs be a \pvari and consider finitely many languages $L_1,\dots,L_k \in \Cs$. There exists a \Cs-morphism $\eta: A^* \to (N,\leq)$ such that $L_1,\dots,L_k$ are recognized by $\eta$.
\end{proposition}

Finally, we state the following simple lemma, which considers group languages.

\begin{lemma} \label{lem:gmorph}
  Let \Gs be a \vari of group languages and let $\eta: A^* \to G$ be a \Gs-morphism. Then, $G$ is a group.
\end{lemma}

\begin{proof}
  Let $p = \omega(G)$. For every $g \in G$, we write $g\inv = g^{p-1}$. It now suffices to prove that $g^p = gg\inv = g\inv g = 1_G$. By hypothesis, $\alpha\inv(g^p)$ is a group language. Hence, we have a morphism $\beta: A^*\to H$ into a finite group $H$ recognizing $\alpha\inv(1_G)$. Since $\veps \in \alpha\inv(1_G)$, it is immediate that $\beta\inv(1_H) \subseteq \alpha\inv(1_G)$. Let $w \in \alpha\inv(g^p)$ and $n = \omega(H)$. Since $H$ is a finite group, $(\beta(w))^n = 1_H$ (the unique idempotent in $H$). Therefore, $\beta(w^n) = 1_H$, which implies that $w^n \in \alpha\inv(1_G)$. Hence, $\alpha(w^n) = 1_G$. Since $\alpha(w) = g^p$ which is idempotent, we get $g^p = (g^p)^n = \alpha(w^n) = 1_G$.
\end{proof}

\subsection{\Cs-pairs}

Given a \pvari \Cs and a morphism $\alpha: A^*\to M$, we associate a relation on $M$. The definition is taken from~\cite{PZ:generic18}, where it is used to characterize all classes of the form \pol{\Cs} for an arbitrary \pvari \Cs (we recall this characterization below). We say that $(s,t) \in M^2$ is a \emph{\Cs-pair} (for $\alpha$) if and only if $\alpha\inv(s)$ is \emph{not} \Cs-separable from $\alpha\inv(t)$. The \Cs-pair relation is not robust. One may verify that it is reflexive when $\alpha$ is surjective and symmetric when \Cs is closed under complement. However, it is \emph{not} transitive in general. We shall use the following lemma, which connects this notion to \Cs-morphisms.

\begin{lemma}\label{lem:pairsm}
  Let \Cs be a \pvari and let $\alpha: A^* \to M$ be a morphism into a finite monoid. The two following properties hold:
  \begin{itemize}
    \item for every \Cs-morphism $\eta: A^* \to (N,\leq)$ and every \Cs-pair $(s,t) \in M^2$ for $\alpha$, there exist $u,v \in A^*$ such that $\eta(u) \leq \eta(v)$, $\alpha(u) = s$ and $\alpha(v) = t$.
    \item there exists a \Cs-morphism $\eta: A^* \to (N,\leq)$ such that for all $u,v \in A^*$, if $\eta(u) \leq \eta(v)$, then $(\alpha(u),\alpha(v))$ is a \Cs-pair for $\alpha$.
  \end{itemize}
\end{lemma}

\begin{proof}
  For the first assertion, let $\eta: A^* \to (N,\leq)$ be a \Cs-morphism and let $(s,t) \in M^2$ be a \Cs-pair for $\alpha$. Let $F \subseteq N$ be the set of all elements $r \in N$ such that $\eta(u) \leq r$ for some $u \in \alpha\inv(s)$. By definition, $F$ is an upper set for the ordering $\leq$ on $N$. Hence, $\eta\inv(F) \in \Cs$. Moreover, it is immediate from the definition of $F$ that $\alpha\inv(s) \subseteq \eta\inv(F)$. Since $(s,t)$ is a \Cs-pair, it follows that $\eta\inv(F) \cap \alpha\inv(t) \neq \emptyset$. This yields $v \in A^*$ such that $\eta(v) \in F$ and $\alpha(v) = t$. Finally, the definition of $F$ yields $u \in A^*$ such that $\eta(u) \leq \eta(v)$ and $\alpha(u) = s$, concluding the proof.

  For the second assertion, let $P \subseteq M^2$ be the set of all pairs $(s,t) \in M^2$ which are \emph{not} \Cs-pairs. For every $(s,t) \in P$, there exists $K_{s,t} \in \Cs$ separating $\alpha\inv(s)$ from $\alpha\inv(t)$. Proposition~\ref{prop:genocm} yields a \Cs-morphism $\eta: A^* \to (N,\leq)$ such that every language $K_{s,t}$ for $(s,t) \in P$ is recognized by $\eta$. It remains to prove that for every $u,v \in A^*$, if $\eta(u) \leq \eta(v)$, then $(\alpha(u),\alpha(v))$ is a \Cs-pair. We prove the contrapositive. Assuming that $(\alpha(u),\alpha(v))$ is a \emph{not} a \Cs-pair, we prove \mbox{$\eta(u)\not\leq \eta(v)$}. By hypothesis, $(\alpha(u),\alpha(v)) \not\in P$. Hence, $K_{\alpha(u),\alpha(v)} \in \Cs$ is defined and separates $\alpha\inv(\alpha(u))$ from $\alpha\inv(\alpha(v))$. Thus, $u \in K_{\alpha(u),\alpha(v)}$ and $v \not\in K_{\alpha(u),\alpha(v)}$. Since $K_{\alpha(u),\alpha(v)}$ is recognized by $\eta$, this implies $\eta(u) \not\leq \eta(v)$, concluding the proof.
\end{proof}

\noindent
{\bf Application to polynomial closure.} We now recall the characterization of \pol{\Cs} from~\cite{PZ:generic18}.

\begin{theorem} \label{thm:polc}
  Let \Cs be a \pvari and let $\alpha: A^* \to (M,\leq)$ be a morphism. Then, $\alpha$ is a \pol{\Cs}-morphism if and only if the following condition holds:
  \begin{equation} \label{eq:polc}
    s^{\omega+1} \leq s^\omega t s^\omega \quad \text{for every \Cs-pair $(s,t) \in M^2$}.
  \end{equation}
\end{theorem}

By definition, one can compute all \Cs-pairs associated to a morphism provided that \Cs-separation is decidable. Hence, in view of Proposition~\ref{prop:synmemb}, it follows from Theorem~\ref{thm:polc} that when \Cs is a \pvari with decidable separation, membership is decidable for \pol{\Cs}.

An interesting point is that Theorem~\ref{thm:polc} can be simplified in the special case when \Cs is a \vari of group languages \Gs or its \wsuit extension $\Gs^+$. This will be useful later when dealing with \bpol{\Gs} and \bpol{\Gs^+}. We first present a specialized characterization of the \pol{\Gs}-morphisms.

\begin{theorem}\label{thm:polg}
  Let \Gs be a \vari of group languages and let $\alpha: A^* \to (M,\leq)$ be a surjective morphism. Then, $\alpha$ is a \pol{\Gs}-morphism if and only if the following condition holds:
  \begin{equation} \label{eq:polg}
    1_M \leq s \quad \text{for every $s \in M$ such that $(1_M,s)$ is a \Gs-pair}.
  \end{equation}
\end{theorem}

\begin{proof}
  If $\alpha$ is a \pol{\Gs}-morphism, it is immediate from Theorem~\ref{thm:polc} that~\eqref{eq:polg} holds, since $(1_M)^p = 1_M$ for every $p \in \nat$. We turn to the converse implication: assume that~\eqref{eq:polg} holds. We show that $\alpha$ is a \pol{\Gs}-morphism. By Theorem~\ref{thm:polc}, it suffices to prove that~\eqref{eq:polc} holds: given a \Gs-pair $(s,t) \in M^2$, we show that $s^{\omega+1} \leq s^\omega t s^\omega$. By Lemma~\ref{lem:pairsm}, there exists a \Gs-morphism $\eta: A^* \to G$ such that for all $u,v \in A^*$, if $\eta(u) = \eta(v)$, then $(\alpha(u),\alpha(v))$ is a \Gs-pair. Moreover, $G$ is a group by Lemma~\ref{lem:gmorph}. Since $(s,t)$ is a \Gs-pair, Lemma~\ref{lem:pairsm} yields $x,y \in A^*$ such that $\eta(x) = \eta(y)$, $\alpha(x) = s$ and $\alpha(y) = t$. Let $n = \omega(G) \times \omega(M)$. Since $G$ is a group, we have $\eta(yx^{n-1}) = \eta(x^n) = 1_G = \eta(\veps)$. Thus, $(1_M,ts^{n-1})$ is a \Gs-pair by definition of $\eta$.  Hence, we get $1_M \leq ts^{n-1}$ from~\eqref{eq:polg}. We may now multiply by $s^n$ on the left and $s$ on the right to get $s^{n+1} \leq s^nts^{n}$. Since $n$ is a multiple of $\omega(M)$, we get $s^{\omega+1} \leq s^\omega t s^\omega$, as desired.
\end{proof}

Finally, we present a similar statement for classes of the form \pol{\Gs^+}.

\begin{theorem} \label{thm:polgp}
  Let \Gs be a \vari of group languages, $\alpha: A^* \to (M,\leq)$ be a surjective morphism and $S = \alpha(A^+)$. Then, $\alpha$ is a \pol{\Gs^+}-morphism if and only if the following condition~holds:
  \begin{equation} \label{eq:polgp}
    e \leq ese \quad \text{for every $e \in E(S)$ and $s \in M$ such that $(1_M,s)$ is a \Gs-pair}.
  \end{equation}
\end{theorem}

\begin{proof}
  Assume first that $\alpha$ is a \pol{\Gs^+}-morphism. We show that~\eqref{eq:polgp} is satisfied. Let $e \in E(S)$ and $s \in M$ such that $(1_M,s)$ is a \Gs-pair. We show that $(e,se)$ is a $\Gs^+$-pair. Since $e$ is idempotent, it will then be immediate from Theorem~\ref{thm:polc} that $e \leq ese$, as desired. By contradiction, assume that $(e,se)$ is \emph{not} a $\Gs^+$-pair. This yields $K \in \Gs^+$ such that $\alpha\inv(e) \subseteq K$ and $K \cap \alpha\inv(se) = \emptyset$. By definition of $\Gs^+$, there exists $L \in \Gs$ such that either $K = \{\veps\} \cup L$ or $K = A^+ \cup L$. Proposition~\ref{prop:genocm} yields a \Gs-morphism $\eta: A^* \to G$ recognizing $L$. Since $(1_M,s)$ is a \Gs-pair, Lemma~\ref{lem:pairsm} yields $x,y \in A^*$ such that $\eta(x) = \eta(y)$, $\alpha(x) = 1_M$ and $\alpha(y) = s$. Moreover, let $u \in A^+$ be a nonempty word such that $\alpha(u) = e$ ($u$ exists since $e \in S = \alpha(A^+)$). Since $\alpha(x) = 1_M$, we have $\alpha(xu) = e$. Hence, we get $xu \in K$ by hypothesis on $K$. This yields $xu \in L$ by definition of $L$ since $xu \in A^+$. Finally, since $\eta(xu) = \eta(yu)$ and $L$ is recognized by $\eta$, we get $yu \in L$. Since $yu \in A^+$, this yields $yu \in K$. This is a contradiction since $\alpha(yu) = se$ and $K \cap \alpha\inv(se) = \emptyset$.

  We turn to the converse implication. Assume that~\eqref{eq:polgp} holds. We show that $\alpha$ is a \pol{\Gs^+}-morphism. By Theorem~\ref{thm:polc}, it suffices to prove that~\eqref{eq:polc} holds for $\Cs = \Gs^+$: given a $\Gs^+$-pair $(s,t) \in M^2$, we show that $s^{\omega+1} \leq s^\omega t s^\omega$. We consider two cases. First, we assume that $s \not\in \alpha(A^+)$. By definition, this exactly says that $s = 1_M$ and $\alpha\inv(1_M) = \{\veps\}$. Since $\{\veps\} \in \Gs^+$, the hypothesis that $(1_M,t) = (s,t)$ is a $\Gs^+$-pair implies that $\{\veps\} \cap \alpha\inv(t) \neq \emptyset$. In other words, we have $s = t = 1_M$ and it is clear that $s^{\omega+1} = s^\omega t s^\omega = 1_M$. We now assume that $s \in \alpha(A^+)$. Clearly, this implies that $s^{\omega} \in E(S)$. Since $\Gs \subseteq \Gs^+$, it is immediate that $(s,t)$ is a \Gs-pair. One may now use the argument from the proof of Theorem~\ref{thm:polg} to obtain that $(1_M,ts^{\omega-1})$ is a \Gs-pair. Hence, it follows from~\eqref{eq:polgp} that $s^\omega \leq s^\omega ts^{2\omega-1}$. It now suffices to multiply by $s$ on the right to get $s^{\omega+1} \leq s^\omega t s^\omega$ as desired.
\end{proof}

\section{Group languages}
\label{sec:bpolg}
In this section, we look at classes of the form \bpol{\Gs} when \Gs is a \vari of group languages. We present a generic algebraic characterization of such classes, which implies that \bpol{\Gs}-membership is decidable when this is already the case for \Gs-separation.

\subsection{Preliminaries}

We present a result that we shall use whenever we need to build a \pol{\Gs}-cover. Let $L \subseteq A^*$ be a language. For every word $w \in A^*$, we associate a language $\uclos_L w \subseteq A^*$. Let $a_1,\dots,a_n \in A$ be the letters such that $w = a_1 \cdots  a_n$. We define $\uclos_L w = La_1L \cdots a_nL \subseteq A^*$ (in particular, we let $\uclos_L \veps = L$). We may now present the statement.

\begin{proposition} \label{prop:pgcov}
  Let $H \subseteq A^*$ be an arbitrary language and let $L \subseteq A^*$ be a group language such that $\veps \in L$. There exists a cover \Kb of $H$ such that every $K \in \Kb$ is of the form $K = \uclos_L w$ for some word $w \in H$.
\end{proposition}

\begin{proof}
  Since $L$ is a group language, it is recognized by a morphism $\eta: A^* \to G$ where $G$ is a finite group. Let $L' = \eta\inv(1_G)$. Clearly, $L'$ is a group language such that $\veps\in L'$ and since $\veps \in L$, we have $L' \subseteq L$.

  We use $L'$ to define an ordering ``$\preceq$'' on $A^*$. Consider two words $u,v \in A^*$, we write $u \preceq v$ when $v \in \uclos_{L'} u$. Since $L' = \eta\inv(1_G)$, is is straightforward to verify that $\eta(u)=\eta(v)$ for every $u,v \in A^*$ such that $u \preceq v$. Since $\veps \in L'$, it~is simple to verify that $\preceq$ is reflexive and antisymmetric. We prove that it is transitive. Let $u,v,w \in A^*$ such that $u \preceq v$ and $v \preceq w$. We show that $u \preceq w$. By definition, we have $v \in \uclos_{L'} u$. Hence, we get $a_1,\dots,a_n \in A$ and $x_0,\dots,x_n \in L' = \eta\inv(1_G)$ such that $u = a_1 \cdots a_n$ and $v = x_0a_1x_1 \cdots a_nx_n$. Since we also have $w \in \uclos_{L'} v$, one may verify that this yields $y_0,\dots,y_n \in A^*$ such that $w = y_0a_1y_1 \cdots a_ny_n$ and $y_i \in \uclos_{L'} x_i$ for every $i \leq n$. The latter property implies that $\eta(y_i) = \eta(x_i) = 1_G$ for every $i \leq n$. Therefore, $y_0,\dots,y_n  \in L'$. We conclude that $w \in \uclos_{L'} u$ which exactly says that $u \preceq w$ as desired. The following lemma states that $\preceq$ is a ``well quasi-order''. A proof is available in~\cite[Proposition 3.10]{cano:hal-01247172}. This can also be shown using a simple generalization of the proof of Higman's lemma.

  \begin{lemma}\label{lem:half:higman}
    Consider an infinite sequence $(u_i)_{i \in \nat}$ of words in $A^*$. There exist $i,j \in \nat$ such that $i<j$ and $u_i \preceq u_j$.
  \end{lemma}

  We may now complete the proof and build the desired cover of the language $H \subseteq A^*$.  We say that a word $v \in H$ is \emph{minimal} if there exists no other word $u \in H$ such that $u \preceq v$. Moreover, we define $F \subseteq H$ as the set of all minimal words of $H$. By definition, we have $u \not\preceq u'$ for every $u,u' \in F$ such that $u \neq u'$.  Hence, it is immediate from Lemma~\ref{lem:half:higman} that $F \subseteq H$ is a finite set. We define $\Kb = \{\uclos_L u \mid u \in F\}$. It remains to prove that \Kb is a cover of $H$. Since \Kb is finite by definition, we have to prove that for every $v \in H$, there exists $u \in F$ such that $v \in \uclos_L u$. We fix $v$ for the proof. If $v$ is minimal, then $v \in F$ and it is clear that $v\in\uclos_L v$ since $\veps \in L$. Assume now that $v$ is \emph{not} minimal. In that case, there exists another word $u \in H$ which is minimal and such that $u \preceq v$. Since $u$ is minimal, we have $u \in F$. Thus, it suffices to prove that $v \in \uclos_L u$. Since $u \preceq v$, we have $v \in \uclos_{L'} u$ by definition. Moreover, since $L' \subseteq L$, it is immediate that $\uclos_{L'} u \subseteq \uclos_{L} u$. Consequently, we obtain that $v \in \uclos_L u$, which completes the proof.
\end{proof}

\subsection{Characterization of \bpol\Gs}

We are ready to present the characterization. As announced, we actually characterize the \bpol{\Gs}-morphisms. Recall that since \bpol{\Gs} is a \vari, it suffices to consider unordered monoids by Lemma~\ref{lem:cmorphbool}.

\begin{theorem} \label{thm:bpolg}
  Let \Gs be a \vari of group languages and let $\alpha: A^* \to M$ be a surjective morphism. Then, $\alpha$ is a $\bpol{\Gs}$-morphism if and only if the following condition holds:
  \begin{equation}\label{eq:gone}
    \begin{array}{c}
      (qr)^{\omega}(st)^{\omega+1} = (qr)^{\omega}qt(st)^{\omega} \\
      \text{for every $q,r,s,t \in M$ such that $(q,s)$ is a \Gs-pair.}
    \end{array}
  \end{equation}
\end{theorem}

Computing the \Gs-pairs associated to a morphism boils down to \Gs-separation. Hence, in view of Proposition~\ref{prop:synmemb}, Theorem~\ref{thm:bpolg} implies that if \emph{separation} is decidable for a \vari of group languages~\Gs, then \emph{membership} is decidable for~\bpol{\Gs}.

\begin{remark}
  The decidability result itself is not new. In fact, it is even known~\cite{pzconcagroup} that \emph{separation} is decidable for \bpol{\Gs} when this is already the case for \Gs. Our main contribution is the algebraic characterization and its proof, which relies on self-contained language theoretic arguments.
\end{remark}

We can also use Theorem~\ref{thm:bpolg} in order to reprove well-known results for particular classes \Gs. For example, since $\stzer = \{\emptyset,A^*\}$, every pair $(s,t) \in M^2$ is an \stzer-pair. Hence, using Theorem~\ref{thm:bpolg}, one may verify that a surjective morphism $\alpha: A^*\to M$ is a \bpol{\stzer}-morphism if and only if the equation $(st)^\omega s=(st)^\omega=t(st)^\omega$ holds for every $s,t \in M$. This is exactly the characterization of the class $\bpol{\stzer} = \bsc{1}(<)$ of piecewise testable languages by Simon~\cite{simonthm}. We also get a characterization of the class $\bpol{\md} = \bsc{1}(<,\mathit{MOD})$. Though the statement does not really simplify in this case, it is easily shown to be equivalent to the one presented in~\cite{ChaubardPS06}. Finally, there exists a simple characterization of \bpol{\grp} presented in~\cite{henckell:hal-00019815}: a surjective morphism $\alpha: A^*\to M$ is a \bpol{\grp}-morphism if and only if $(ef)^\omega = (fe)^\omega$ for all idempotents $e,f \in E(M)$. This is also a corollary of Theorem~\ref{thm:bpolg}. Yet, this requires a bit of technical work as well as a knowledge of the \grp-separation algorithm~\cite{Ash91} (which one needs for describing the \grp-pairs).

\begin{proof}[Proof of Theorem~\ref{thm:bpolg}]
  We first assume that $\alpha$ is a $\bpol{\Gs}$-morphism and prove that it satisfies~\eqref{eq:gone}. There exists a finite set \Hb of languages in \pol{\Gs} such that for every $s \in M$, the language $\alpha\inv(s)$ is a Boolean combination of languages in \Hb. Since \pol{\Gs} is a \pvari, Proposition~\ref{prop:genocm} yields a \pol{\Gs}-morphism $\eta: A^* \to (N,\leq)$ recognizing every $H \in \Hb$. Moreover, Lemma~\ref{lem:pairsm} yields a \Gs-morphism $\beta: A^* \to G$ such that for every $u,v \in A^*$, if $\beta(u) = \beta(v)$, then $(\eta(u),\eta(v)) \in N^2$ is a \Gs-pair for $\eta$. We know that $G$ is a group by Lemma~\ref{lem:gmorph}. We let $n = \omega(M) \times \omega(N) \times \omega(G)$.

  We may now prove that~\eqref{eq:gone} holds. Let $q,r,s,t \in M$ such that $(q,s)$ is a \Gs-pair. We prove that $(qr)^{\omega}(st)^{\omega+1} = (qr)^{\omega}qt(st)^{\omega}$. Since $\beta: A^* \to G$ is a \Gs-morphism and $(q,s)$ is a \Gs-pair, Lemma~\ref{lem:pairsm} yields $u,x \in A^*$ and $g \in G$ such that $\beta(u) = \beta(x) = g$, $\alpha(u) = q$ and $\alpha(x) = s$. Since $\alpha$ is surjective, we get $v,y \in A^*$ such that $\alpha(v) = r$ and $\alpha(y) = t$. Since $G$ is a group, we have $\beta((uv)^n) = \beta((xy)^n) = 1_G$ by definition of $n$.  Let \mbox{$v' = v(uv)^{n-1}$} and  \mbox{$y'=y(xy)^{n-1}$}. Since \mbox{$\beta(u) = \beta(x) = g$}, we get $\beta(v') = \beta(y') = g\inv$, $\beta(uy') = 1_G$ and $\beta(v'x) = 1_G$. Hence, by definition of $\beta$, $(1_N,\eta(uy'))$ and $(1_N,\eta(v'x))$ are \Gs-pairs. Since $\eta$ is a \pol{\Gs}-morphism, Theorem~\ref{thm:polg} yields \mbox{$1_N \leq \eta(uy')$} and \mbox{$1_N \leq \eta(v'x)$}. We may now multiply to obtain \mbox{$\eta((uv)^n (xy)^{n+1}) \leq \eta((uv)^n uy'(xy)^{n+1})$} and \mbox{$\eta((uv)^n uy (xy)^{n}) \leq \eta((uv)^n uv'xy(xy)^{n})$}. By definition of $n$, $y'$ and $v'$, one may verify that this implies that $\eta((uv)^n (xy)^{n+1}) \leq \eta((uv)^n uy(xy)^{n})$ and $\eta((uv)^n uy (xy)^{n}) \leq \eta((uv)^n (xy)^{n+1})$. Altogether, we get $\eta((uv)^n (xy)^{n+1}) = \eta((uv)^n uy(xy)^{n})$. Moreover, since $\eta$ recognizes all $H \in \Hb$ by definition, it follows that $(uv)^n (xy)^{n+1} \in H \Leftrightarrow  (uv)^n uy(xy)^{n} \in H$ for every $H \in \Hb$. Since all languages recognized by $\alpha$ are Boolean combination of languages in~\Hb, we get $\alpha((uv)^n (xy)^{n+1}) = \alpha((uv)^n uy(xy)^{n})$. By definition, this exactly says that $(qr)^{\omega}(st)^{\omega+1} = (qr)^{\omega}qt(st)^{\omega}$ as desired.

  \smallskip

  We turn to the converse implication. Assume that $\alpha$ satisfies~\eqref{eq:gone}. We prove that $\alpha$ is a \bpol{\Gs}-morphism. Lemma~\ref{lem:pairsm} yields a \Gs-morphism $\beta: A^* \to G$ such that for every $u,v \in A^*$, if $\beta(u) = \beta(v)$, then $(\alpha(u),\alpha(v))$ is a \Gs-pair. We write $L = \beta\inv(1_G) \in \Gs$. By hypothesis on \Gs, $L$ is a group language. Moreover, we have $\veps \in L$ by definition. Given a finite set of languages \Kb, and $s,t \in M$, we say that \Kb is \emph{$(s,t)$-safe} if for every $K \in \Kb$ and $w,w' \in K$, we have $s\alpha(w)t = s\alpha(w')t$. The argument is based on the following lemma.

  \begin{lemma}\label{lem:bpgmain}
    Let $s,t \in M$. There exists a \bpol{\Gs}-cover of $L$ which is $(s,t)$-safe.
  \end{lemma}

  Before proving Lemma~\ref{lem:bpgmain} we first use it to prove that every language recognized by $\alpha$ belongs to \bpol{\Gs}, thus concluding the argument. We apply Lemma~\ref{lem:bpgmain} with $s = t=1_M$. This yields a \bpol{\Gs}-cover $\Kb_L$ of $L$ which is $(1_M,1_M)$-safe. We use it to build a \bpol{\Gs}-cover \Kb of $A^*$ which is $(1_M,1_M)$-safe. Since $L \in \Gs$ and $\veps \in L$, Proposition~\ref{prop:pgcov} yields a cover \Pb of $A^*$ such that every $P \in \Pb$, there exist $n \in \nat$ and $a_1,\dots,a_n \in A$ such that $P = La_1L \cdots a_nL$. We cover each $P \in \Pb$ independently. Consider a language $P \in \Pb$. By definition, $P = La_1L \cdots a_nL$ for $a_1,\dots,a_n \in A$. Since $L \in \Gs$ and $\Kb_L$ is a \bpol{\Gs}-cover of $L$,  Proposition~\ref{prop:bconcat} yields a \bpol{\Gs}-cover $\Kb_P$ of $P =  La_1L \cdots a_nL$ such that for every $K \in \Kb_P$, there exist $K_0,\dots,K_n \in \Kb_L$ satisfying $K \subseteq K_0a_1K_1 \cdots a_nK_n$. Since $\Kb_L$ is $(1_M,1_M)$-safe, it is immediate that $\Kb_P$ is $(1_M,1_M)$-safe as well. Finally, since \Pb is a cover of $A^*$, it is now immediate that $\Kb=\bigcup_{P\in\Pb}\Kb_P$ is a $(1_M,1_M)$-safe \bpol{\Gs}-cover of $A^*$. Since \Kb is $(1_M,1_M)$-safe, we know that for every $K \in \Kb$, there exists $s\in M$ such that $K \subseteq \alpha\inv(s)$.  Hence, since \Kb is a cover of $A^*$, it is immediate that for every $F \subseteq M$, the language $\alpha\inv(F)$ is a union of languages in \Kb. By closure under union, it follows that $\alpha\inv(F) \in \bpol{\Gs}$. This exactly says that all languages recognized by $\alpha$ belong to \bpol{\Gs}.

  \smallskip

  It remains to prove Lemma~\ref{lem:bpgmain}. We define a preorder on $M^2$ that we shall use as an induction parameter. Consider $(s,t),(s',t') \in M^2$. We write $(s,t) \leqslant_{L} (s',t')$ if there exist $x,y \in A^*$ such that $xy \in L$, $s' = s\alpha(x)$ and $t' = \alpha(y)t$. It is immediate that $\leqslant_L$ is reflexive since we have $\veps = \veps \veps \in L$. Let us verify that $\leqslant_L$ is transitive. Let $(s,t),(s',t'),(s'',t'') \in M^2$ such that $(s,t) \leqslant_{L} (s',t')$ and $(s',t')\leqslant_{L} (s'',t'')$. We show that $(s,t) \leqslant_{L} (s'',t'')$. By definition, we have $xy,x'y' \in L$ such that $s' =s\alpha(x)$, $t' = \alpha(y)t$, $s'' = s'\alpha(x')$ and $t'' = \alpha(y')t'$. Hence, $s'' = s\alpha(xx')$ and $t'' = \alpha(y'y)t$. Moreover, since $L = \beta\inv(1_G)$, we have $\beta(xx'y'y) = \beta(xy) = 1_G$, which yields $xx'y'y \in L$. We conclude that $(s,t) \leqslant_{L} (s'',t'')$, as desired.

  We may now start the proof. Let $s,t \in M$. We construct a \bpol{\Gs}-cover \Kb of $L$ which is $(s,t)$-safe. We proceed by descending induction on the number of pairs $(s',t') \in M^2$ such that $(s,t) \leqslant_L (s',t')$. We handle the base case and the inductive step simultaneously. Consider a word $w \in L$. We say  \emph{$w$ stabilizes $(s,t)$} if there exist $u,v \in A^*$ such that $uv \in \uclos_L w$, $s\alpha(u) = s$ and $\alpha(v)t = t$. Observe that by definition, \veps stabilizes $(s,t)$ since we have $\veps\veps = \veps \in L = \uclos_L \veps$. We let $H \subseteq L$ be the language of all words $w \in L$ that do \emph{not} stabilize $(s,t)$. Note that by definition~$\veps\not\in H$. We first use induction to build a \bpol{\Gs}-cover $\Kb_H$ of $H$ and then complete it to build \Kb. Let us point out that it may happen that $H$ is empty. This is the base case, it suffices to define~$\Kb_H = \emptyset$.

  \smallskip

  Let $P \subseteq M^2$ be the set of all pairs $(s',t') \in M^2$ such that $(s,t) \leqslant_L (s',t')$ and \mbox{$(s',t') \not\leqslant_{L} (s,t)$}. We define $\ell = |P|$ and write $P = \{(s'_1,t'_1), \dots, (s'_\ell,t'_\ell)\}$. For every $i \leq \ell$, we may apply induction in the proof of Lemma~\ref{lem:bpgmain} to obtain a \bpol{\Gs}-cover $\Kb_i$ of $L$ which is $(s'_i,t'_i)$-safe. We define $\Kb_L = \left\{L \cap K_1 \cap \cdots \cap K_\ell \mid \text{$K_i \in \Kb_{i}$ for every $i \leq \ell$} \right\}$. Since $L \in \Gs$, it is immediate  that $\Kb_L$ is a \bpol{\Gs}-cover of $L$ which is $(s',t')$-safe for every $(s',t')\in P$. We use it to construct $\Kb_H$.

  \begin{lemma}\label{lem:ginduc}
    There exists an $(s,t)$-safe \bpol{\Gs}-cover $\Kb_H$ of $H$.
  \end{lemma}

  \begin{proof}
    Since $L$ is a group language such that $\veps \in L$, Proposition~\ref{prop:pgcov} yields a cover \Ub of $H$ such that for every $U \in \Ub$, there exist $n \geq 1$ and $a_1,\dots,a_n \in A$ such that $a_1 \cdots a_n \in H$ and $U = La_1L \cdots a_nL$ (note that $n \geq 1$ as $\veps\not\in H$). For each $U \in \Ub$, we build an $(s,t)$-safe \bpol{\Gs}-cover $\Kb_U$ of $U$. Since $\Ub$ is a cover of $H$, it will then suffice to define $\Kb_H$ as the union of all covers $\Kb_U$.  We fix $U \in \Ub$.

    By definition, $U = La_1L \cdots a_nL$ where $a_1 \cdots a_n \in H$. Since $L \in \Gs$, $\veps \in L$ and $\Kb_L$ is a \bpol{\Gs}-cover of $L$, Proposition~\ref{prop:bconcat} yields a \bpol{\Gs}-cover $\Kb_U$ of $U$ such that for each $K \in \Kb_U$, we have $K \subseteq K_0a_1K_1 \cdots a_nK_n$ for $K_0,\dots,K_n \in \Kb_L$. It remains to show that $\Kb_U$ is $(s,t)$-safe. We fix $K \in \Kb_U$ as described above and $w,w' \in K$. We show that $s\alpha(w)t = s\alpha(w')t$.  By definition, we have \mbox{$w_i,w'_i \in K_i$} for all $i \leq n$ such that $w = w_0a_1w_1 \cdots a_nw_n$ and $w' = w'_0a_1w'_1 \cdots a_nw'_n$. We let $u_i = w_0a_1 \cdots w_{i-1}a_i$ and $u'_i = w'_0a_1 \cdots w'_{i-1}a_i$  for $0  \leq i \leq n$ ($u_0 = u'_0 = \veps$). We also let $v_i = a_{i+1}w_{i+1} \cdots a_nw_n$ and $v'_i = a_{i+1}w'_{i+1} \cdots a_nw'_n$ ($v_n= v'_n = \veps$). Note that $u_iw'_iv'_i = u_{i-1}w_{i-1}v'_{i-1}$ for $1\leq i \leq n$. Hence, it suffices to prove that $s\alpha(u_iw_iv'_i)t = s\alpha(u_iw'_iv'_i)t$ for $0 \leq i \leq n$. By transitivity, it will then follow that $s\alpha(u_nw_nv'_n)t = s\alpha(u_0w'_0v'_0)t$, \emph{i.e.}, $s\alpha(w)t = s\alpha(w')t$ as desired.

    We fix $i\leq n$ and show that $s\alpha(u_iw_iv'_i)t = s\alpha(u_iw'_iv'_i)t$.  By hypothesis, $w_i,w'_i \in K_i$. Since $K_i \in \Kb_L$  is $(s',t')$-safe for all $(s',t')\in P$, it suffices to prove that $(s\alpha(u_i),\alpha(v'_i)t) \in P$. There are two conditions to verify. First, we show that $(s,t) \leqslant_L (s\alpha(u_i),\alpha(v'_i)t)$. By definition of $\leqslant_L$, this boils down to proving that $u^{}_iv'_i \in L$. By definition, $w_j,w'_j \in K_j$ for every $j \leq n$. Moreover, since $K_j \in \Kb_L$, it follows that $w_j,w'_j \in L$ for every $j \leq n$ by definition of $\Kb_L$. It follows that $\beta(w_j) = \beta(w'_j) = 1_G$ since $L = \beta\inv(1_G)$. Therefore, by definition of $u_i$ and $v'_i$, we obtain $\beta(u_i) = \beta(a_1 \cdots a_i)$ and $\beta(v'_i) = \beta(a_{i+1} \cdots a_n)$. This yields $\beta(u_iv'_i) = \beta(a_1 \cdots a_n)$. Finally, since $a_1 \cdots a_n \in H \subseteq L$ and $L$ is recognized by $\beta$, we get $u^{}_i v'_i \in L$, as desired. It remains to prove that $(s\alpha(u_i),\alpha(v'_i)t) \not\leqslant_{L} (s,t)$. By contradiction, assume that $(s\alpha(u_i),\alpha(v'_i)t) \leqslant_{L} (s,t)$. This yields $x,y \in A^*$ such that $xy \in L$ and $s = s\alpha(u_ix)$ and $t = \alpha(yv'_i)t$. Since $xy \in L$ and $w_j,w'_j \in L$, it is immediate by definition of $u_i$ and $v'_i$ that $u_ixyv_i \in \uclos_L (a_1 \cdots a_n)$. Hence,  $a_1 \cdots a_n$ stabilizes $(s,t)$. This is a contradiction since $a_1 \cdots a_n \in H$.
  \end{proof}

  We are ready to construct the desired $(s,t)$-safe \bpol{\Gs}-cover \Kb of $L$. Let $\Kb_H$ be the \bpol{\Gs}-cover of $H$ given by Lemma~\ref{lem:ginduc}. We let $K_\bot = L \setminus (\bigcup_{K \in \Kb_H} K)$. Finally, we define $\Kb = \{K_\bot\} \cup \Kb_H$. It is immediate that \Kb is a \bpol{\Gs}-cover of $L$ since \bpol{\Gs} is a Boolean algebra (recall that $L \in \Gs$). It remains to verify that \Kb is $(s,t)$-safe. Since we already know that $\Kb_H$ is $(s,t)$-safe, it suffices to prove that for every $w,w' \in K_{\bot}$, we have $s\alpha(w)t = s\alpha(w')t$. We actually show that $s\alpha(w)t = st$ for every $w \in K_\bot$. Since this is immediate when $w = \veps$, we assume that $w \in A^+$ and let $a_1,\dots,a_n \in A$ be the letters such that $w = a_1 \cdots a_n$.

  By definition of $K_\bot$, we know that $w \not\in K'$ for every $K' \in \Kb_H$. Since $\Kb_H$ is a cover of $H$, it follows that $w \not\in H$, which means that $w$ stabilizes $(s,t)$ by definition of $H$. We get $u',v' \in A^*$ such that $u'v' \in \uclos_L w$, $s\alpha(u') = s$ and $\alpha(v')t = t$. Since $u'v' \in \uclos_L w$, there exist $0 \leq i \leq n$ and $x_0,\dots,x_i, y_i,\dots,y_n \in A^*$ which satisfy $x_0,\dots,x_{i-1},x_iy_i, y_{i+1},\dots,y_n \in L$, $u' = x_0a_1x_1 \cdots a_ix_i$ and $v' = y_ia_{i+1}x_{i+1} \cdots a_nx_n$. We write $u = a_1 \cdots a_i$ and $v = a_{i+1} \cdots a_n$. By definition $w = uv$. We show that $s = s \alpha(ux_i)$ and $t = \alpha(y_iv)t$. Let us first assume that this holds and explain why this implies $st = s\alpha(w)t$.

  Since $uv = w$ and $w \in K_\bot \subseteq L = \beta\inv(1_G)$, we have $\beta(u)\beta(v) = 1_G$. Let $p = \omega(G)$. We have $1_G = \beta((y_iv)^p)$. Thus, since $G$ is a group, it follows that $\beta(u) = \beta((y_iv)^{p-1}y_i)$. By definition of $\beta$, it follows that $(\alpha(u),\alpha((y_iv)^{p-1}y_i))$ is a \Gs-pair. Consequently, we obtain from~\eqref{eq:gone} that,
  \[
    (\alpha(ux_i))^\omega (\alpha((y_iv)^{p-1}y_iv))^{\omega+1} = (\alpha(ux_i))^\omega \alpha(uv) (\alpha((y_iv)^{p-1}y_iv))^{\omega}.
  \]
  We may now multiply by $s$ on the left and $t$ on the right. Since $s = s \alpha(ux_i)$ and $t = \alpha(y_iv)t$, this yields $st = s\alpha(uv)t$. This concludes the proof since $uv = w$.

  It remains to show that $s = s \alpha(ux_i)$ and $t = \alpha(y_iv)t$. We prove the former (the latter is symmetrical and left to the reader). For every $j$ such that $0 \leq j \leq i$, we write $z_j = x_ja_{j+1} \cdots x_{i-1}a_{i}x_i$ (when $i = j$, we let $z_i = x_i$). We use induction on $i$ to prove that $s = s \alpha(a_1 \cdots a_{j}z_j)$ for $0 \leq j \leq i$. Clearly, the case $j = i$ yields  $s =s \alpha(a_1 \cdots a_{i}x_i)$ which exactly says that $s = s \alpha(ux_i)$. When $j = 0$, we have $z_0 = x_0a_1x_1\cdots a_ix_i = u'$ and $s\alpha(u') = s$ by hypothesis. Assume now that $1 \leq j \leq i$. Since $x_{j-1} \in L$ and $L = \beta\inv(1_G)$, we have $\beta(x_{j-1}) = \beta(\veps) = 1_G$. Hence, $(\alpha(x_{j-1}),1_M)$ is a \Gs-pair by definition of $\beta$. Applying~\eqref{eq:gone} with the values $\alpha(x_{j-1})$, $\alpha(a_{j}z_{j}a_1\cdots a_{j-1})$, $1_M,1_M$ yields that,
  \begin{equation} \label{eq:mbp:grpeq}
    (\alpha(x_{j-1}a_jz_ja_1 \cdots a_{j-1}))^\omega =  (\alpha(x_{j-1}a_jz_ja_1 \cdots a_{j-1}))^\omega \alpha(x_{j-1}).
  \end{equation}
  By induction hypothesis, we know that $s = s \alpha(a_1 \cdots a_{j-1}z_{j-1})$. Since it is immediate by definition that $a_1 \cdots a_{j-1}z_{j-1} = a_1 \cdots a_{j-1}x_{j-1}a_jz_{j}$, we get,
  \[
    \begin{array}{llll}
      s & = & s \alpha(a_1 \cdots a_{j-1}x_{j-1}a_jz_{j}) &\\
        & = & s (\alpha(a_1 \cdots a_{j-1}x_{j-1}a_jz_{j}))^{\omega+1}        &                                                \\
        & = & s \alpha(a_1 \cdots a_{j-1}) (\alpha(x_{j-1}a_jz_ja_1 \cdots a_{j-1}))^\omega \alpha(x_{j-1})\alpha(a_jz_j)& \\
        & =  & s \alpha(a_1 \cdots a_{j-1}) (\alpha(x_{j-1}a_jz_ja_1 \cdots a_{j-1}))^\omega \alpha(a_jz_j) &  \text{by~\eqref{eq:mbp:grpeq}}\\
        & = & s (\alpha(a_1 \cdots a_{j-1}x_{j-1}a_jz_{j}))^{\omega} \alpha(a_1 \cdots a_{j-1}a_jz_{j})& \\
        & = & s \alpha(a_1 \cdots a_{j}z_j).&
    \end{array}
  \]
  This concludes the proof.
\end{proof}

\section{Well-suited extensions}
\label{sec:bpolgp}
We now consider the classes \bpol{\Gs^+} where \Gs is an arbitrary \vari of group languages. In this case as well, we present a generic algebraic characterization, which implies that \bpol{\Gs^+}-membership is decidable when this is already the case for \Gs-separation. Again, we start with a preliminary result that we shall use to build \pol{\Gs^+}-covers in the proof.

\subsection{Preliminaries}

Consider an arbitrary morphism $\alpha: A^* \to M$ and a \emph{nonempty} word $w \in A^+$. An \emph{$\alpha$-guarded decomposition of $w$} is a tuple $(w_1,\dots, w_{n+1})$ for some $n \in \nat$ such that $w_1,\dots,w_{n+1} \in A^+$ are nonempty words, $w = w_1 \cdots w_{n+1}$ and, if $n \geq 1$, then for every $i$ satisfying $1 \leq i \leq n$, there exists an \emph{idempotent} $e_i \in \alpha(A^+)$ such that $\alpha(w_{i})e_i = \alpha(w_{i})$ and $e_i\alpha(w_{i+1}) = \alpha(w_{i+1})$. We may now present the statement. We prove it as a corollary of Proposition~\ref{prop:pgcov}.

\begin{proposition} \label{prop:pgpcov}
  Let $H \subseteq A^+$ be a language, $\alpha: A^* \to M$ be a morphism and $L \subseteq A^*$ be a group language such that $\veps \in L$. There exists a cover \Kb of $H$ such that for each $K \in \Kb$, there exist a word $w \in H$ and an $\alpha$-guarded decomposition $(w_1,\dots, w_{n+1})$ of $w$ for some $n \in \nat$ such that  $K = w_1 L \cdots w_nLw_{n+1}$ (if $n=0$, then $K=\{w_1\}$).
\end{proposition}

\begin{proof}
  We fix $k = |M|^2$ for the proof. We define an auxiliary alphabet \frB. Intuitively, we use the words in $\frB^+$ to represent the $\alpha$-guarded decompositions of any word in $A^+$ whose length is strictly greater than $k$. We write $E \subseteq \alpha(A^+)$ for the set of all idempotents in $\alpha(A^+)$. Consider the following sets:
  \[
    \begin{array}{lll}
      \frB_\ell & = & \{(w,f) \in A^+ \times E \mid |w| \leq 2k \text{ and } \alpha(w)f = \alpha(w)\}. \\
      \frB_c & = & \{(e,w,f) \in E \times A^+ \times E \mid |w| \leq 2k \text{ and } e\alpha(w)f = \alpha(w)\}. \\
      \frB_r & = & \{(e,w) \in E \times A^+ \mid |w| \leq k \text{ and } e\alpha(w) = \alpha(w)\}.
    \end{array}
  \]
  We define $\frB = \frB_\ell \cup \frB_r \cup \frB_c$. It is clear from the definition that \frB is finite. We use it as an alphabet and define a morphism $\gamma: \frB^* \to A^*$. Let $b \in \frB$. There exists a nonempty word $w \in A^+$ and $e,f \in E$ such that $b = (w,f) \in \frB_\ell$, $b = (e,w) \in \frB_r$ or $b = (e,w,f) \in \frB_c$. We define $\gamma(b) = w$. Moreover, we write $\gamma_c: \frB_c^* \to A^*$ for the restriction of $\gamma$ to $\frB_c^*$. Finally, we say that a word $x \in \frB^*$ is \emph{well-formed} if $x \in \frB_\ell \frB_c^* \frB_r$ (in particular, $|x| \geq 2$) and $x$ is of the form $x = (w_1,f_1)(e_2,w_2,f_2) \cdots (e_n,w_n,f_n)(e_{n+1},w_{n+1})$ where $f_i = e_{i+1}$ for every $i \leq n$. The following lemma can be verified from the definitions.

  \begin{lemma} \label{lem:wformg}
    Let $b_1,\dots,b_m \in \frB$ be letters such that the word $x = b_1 \cdots b_m \in \frB^+$ is well-formed. Then, $(\gamma(b_1),\dots,\gamma(b_m))$ is an $\alpha$-guarded decomposition of the word $\gamma(x) \in A^+$.
  \end{lemma}

  Intuitively Lemma~\ref{lem:wformg} states that every well-formed word in $x \in \frB^+$ encodes an  $\alpha$-guarded decomposition of some word in $A^+$. We handle the converse direction in the following lemma: for every word $w \in A^+$ of sufficient length, there exists an $\alpha$-guarded decomposition of $w$ which is encoded by a word in $\frB^+$.

  \begin{lemma} \label{lem:half:wformed}
    For every $w \in A^+$ such that $|w| > k$, there exists $x \in \frB^+$ which is well-formed and such that $w = \gamma(x)$.
  \end{lemma}

  \begin{proof}
    We proceed by induction on the length of $w$. Since $|w| > k$, there exist $a_0,\dots,a_{k} \in A$ and $w'\in A^*$ such that $w = w'a_0 \cdots a_k$. Since $k = |M|^2$, we may apply the pigeon-hole principle to obtain $i,j$ such that $0 \leq i < j \leq k$, $\alpha(a_0\cdots a_{i}) =  \alpha(a_0\cdots a_{j})$ and $\alpha(a_{i+1} \cdots a_{k}) =  \alpha(a_{j+1}\cdots a_{k})$. Let $u = a_0\cdots a_{i}$ and $v = a_{i+1} \cdots a_{k}$. We have $u,v \in A^+$, $|u|\leq k$ and $|v| \leq k$. Moreover, $w = w'uv$. We consider the idempotent $e = (\alpha(a_{i+1} \cdots a_j))^\omega \in E$. By definition, we have $\alpha(u)e = \alpha(u)$ and $e\alpha(v) = \alpha(v)$. There are now two cases depending on $w'$.

    Assume first that $|w'| \leq k$. In that case $|w'u| \leq 2k$ which implies that $(w'u,e) \in \frB_\ell$ since $\alpha(u)e = \alpha(u)$. Moreover, we have $(e,v) \in \frB_r$ since $|v| \leq k$ and $e\alpha(v) = \alpha(v)$. Consequently, $x = (w'u,e)(e,v) \in \frB^+$ is a well-formed word such that $\gamma(x) = w'uv = w$. Assume now that $|w'| > k$. Since it is clear that $|w'| < |w|$, induction yields a well-formed word $x' \in \frB^+$ such that $\gamma(x') = w'$. By definition $x' = x''(f,v')$ where $x'' \in \frB^+$ and $(f,v') \in \frB_r$. In particular, we have $|v'| \leq k$ and $f\alpha(v') = \alpha(v')$ by definition of $\frB_r$. Hence, $|v'u| \leq 2k$ which implies that $(f,v'u,e) \in \frB_c$ since $\alpha(u)e = \alpha(u)$. Moreover, we have $(e,v) \in \frB_r$ since $|v| \leq k$ and $e\alpha(v) = \alpha(v)$. Let $x = x''(f,v'u,e)(e,v)$. Clearly, $x$ is well-formed since $x' = x''(f,v)$ is. Moreover, $\gamma(x) = \gamma(x''(f,v'))uv = w'uv = w$. This concludes the proof.
  \end{proof}

  We now prove Proposition~\ref{prop:pgpcov}.  We define $L_c=\gamma_c\inv(L) \subseteq \frB_c^*$. Since $L$ is a group language (over $A$) and $\veps \in L$, one may verify that $L_c$ is also a group language (over $\frB_c$) and $\veps\in L_c$. Let $b_\ell \in \frB_\ell$ and $b_r \in \frB_r$. We define,
  \[
    H_{b_\ell,b_r} = \{x \in \frB_c^* \mid \text{$b_\ell x b_r \in \frB^*$ is well-formed and $\gamma(b_\ell x b_r ) \in H$}\}.
  \]
  Proposition~\ref{prop:pgcov} yields a \emph{finite} set $F_{b_\ell,b_r} \subseteq H_{b_\ell,b_r} \subseteq \frB_c^*$ such that $\{\uclos_{L_c} x \mid x \in F_{b_\ell,b_r}\}$ is a cover of $H_{b_\ell,b_r}$. We are ready to build our cover \Kb of $H \subseteq A^*$. For every word $x = b_1 \cdots b_n  \in \frB_c^*$, every  $b_\ell \in \frB_\ell$ and every $b_r \in \frB_r$, we associate the language $[x]_{b_\ell,b_r} = \gamma(b_\ell)L\gamma(b_1)L \cdots \gamma(b_n)L\gamma(b_r) \subseteq A^+$. Finally, we define,
  \[
    \Kb = \{\{w\} \mid w \in H \text{ and } |w| \leq k\} \cup \bigcup_{b_\ell \in \frB_\ell} \bigcup_{b_r \in \frB_r} \{[x]_{b_\ell,b_r} \mid x \in F_{b_\ell,b_r}\}.
  \]
  It remains to prove that \Kb is the desired cover of $H$. First, let us verify that every $K \in \Kb$ is of the form $K = w_1 L \cdots w_nLw_{n+1}$ where $(w_1,\dots, w_{n+1})$ is an $\alpha$-guarded decomposition of some word $w \in H$. This immediate if $K = \{w\}$ for some $w \in H$. We have to handle the case when $K = [x]_{b_\ell,b_r}$ for some $x \in F_{b_\ell,b_r}$. By definition, $x \in H_{b_\ell,b_r}$ which means that $b_\ell x b_r \in \frB^*$ is well-formed and $\gamma(b_\ell x b_r ) \in H$. Let $b_1,\dots,b_n \in \frB_c^*$ be the letters such that $x = b_1 \cdots b_n$. Since $b_\ell b_1 \cdots b_n b_r \in \frB^*$ is well-formed, Lemma~\ref{lem:wformg} yields that $(\gamma(b_\ell),\gamma(b_1),\dots,\gamma(b_n),\gamma(b_r))$ is an $\alpha$-guarded decomposition of $\gamma(b_\ell x b_r ) \in H$. This concludes the proof since $K=  [x]_{b_\ell,b_r} = \gamma(b_\ell)L\gamma(b_1)L \cdots \gamma(b_n)L\gamma(b_r)$.

  We now prove that \Kb is a cover of $H$. It is immediate by definition that \Kb is finite. Given $w \in H$, we exhibit $K \in \Kb$ such that $w \in K$. This is immediate if $|w|\leq k$: we have $\{w\} \in \Kb$ in that case. We now consider the case $|w|>k$. Lemma~\ref{lem:half:wformed} yields $x \in B^+$ which is well-formed and such that $w = \gamma(x)$. By definition of well-formed words $x = b_\ell y b_r$ where $y \in \frB_c^*$, $b_\ell \in \frB_\ell$ and $b_r \in \frB_r$. Therefore, since $\gamma(x) = w \in H$, we have $y \in H_{b_\ell,b_r}$ by definition. Hence, since $\{\uclos_{L_c} z \mid z \in F_{b_\ell,b_r}\}$ is a cover of $H_{b_\ell,b_r}$, we get $z \in  F_{b_\ell,b_r}$ such that $y \in \uclos_{L_c} z$.  We prove that $w \in [z]_{b_\ell,b_r}$ which concludes the proof since $[z]_{b_\ell,b_r} \in \Kb$ by definition. We have $\uclos_{L_c} z = L_cb_1L_c \cdots b_nL_c$ where $b_1,\dots,b_n \in \frB_c$ are the letters such that $b_1 \cdots b_n = z \in H_{b_\ell,b_r}$. Therefore, since $y \in \uclos_{L_c} z$, this yields $x_0, \dots, x_n \in L_c$ such that $y = x_0b_1x_1 \cdots b_nx_n$. Altogether, it follows that $x = b_\ell x_0b_1x_1 \cdots b_nx_n b_r$. Since $w = \gamma(x)$, we get $w = \gamma(b_\ell) \gamma(x_0) \gamma(b_1) \gamma(x_1) \cdots \gamma(b_n) \gamma(x_n) \gamma(b_r)$. Finally, since $L_c = \gamma_c\inv(L)$ and $x_0,\dots,x_n \in L_c$, we have $\gamma(x_i) \in L$ for every $i \leq n$. Hence, we obtain that $w \in \gamma(b_\ell)L\gamma(b_1)L \cdots \gamma(b_n)L\gamma(b_r)$. This exactly says that $w \in [z]_{b_\ell,b_r}$ since $b_1 \cdots b_n = z$ by definition. This concludes the proof.
\end{proof}

\subsection{Characterization}

We may now present the characterization. As we explained, we actually characterize the \bpol{\Gs^+}-morphisms. Recall that since \bpol{\Gs^+} is a \vari, it suffices to consider unordered monoids by Lemma~\ref{lem:cmorphbool}.

\begin{theorem}  \label{thm:wgroup}
  Let \Gs be a \vari of group languages, let $\alpha: A^* \to M$ be a surjective morphism and $S = \alpha(A^+)$. Then, $\alpha$ is a $\bpol{\Gs^+}$-morphism if and only if the following condition holds:
  \begin{equation}\label{eq:wgone}
    \begin{array}{c}
      (eqfre)^{\omega}(esfte)^{\omega+1} = (eqfre)^{\omega}qft(esfte)^{\omega} \\ \text{for all $q,r,s,t \in M$ and $e,f \in E(S)$ such that $(q,s)$ is a \Gs-pair.}
    \end{array}
  \end{equation}
\end{theorem}

Again, by Proposition~\ref{prop:synmemb}, Theorem~\ref{thm:wgroup} implies that if  \emph{separation} is decidable for a \vari of group languages \Gs, then \emph{membership} is decidable for \bpol{\Gs^+}.

Theorem~\ref{thm:wgroup} can also be used to reprove famous results for specific classes~\Gs. As seen in Section~\ref{sec:bpolg}, since $\stzer = \{\emptyset,A^*\}$, every pair $(s,t) \in M^2$ is an \stzer-pair. Hence, one may verify from Theorem~\ref{thm:wgroup} that a surjective morphism \mbox{$\alpha: A^*\to M$} is a \bpol{\stzer^+}-morphism if and only if $(eqfre)^{\omega}(esfte)^{\omega} = (eqfre)^{\omega}qft(esfte)^{\omega}$ for every $q,r,s,t \in S$ and $e,f \in E(S)$ (where $S = \alpha(A^+)$). This is exactly the well-known characterization of the languages of dot-depth one by Knast~\cite{knast83} (\emph{i.e.}, the class $\bpol{\stzer^+} = \bsc{1}(<,+1)$). Additionally, there exists a specialized characterization of $\bpol{\md^+} = \bsc{1}(<,+1,MOD)$ in the literature~\cite{MACIEL2000135}. It can also be reproved as a corollary of Theorem~\ref{thm:wgroup}. However, this requires some technical work involving the \md-pairs.

\begin{proof}[Proof of Theorem~\ref{thm:wgroup}]
  Assume first that $\alpha$ is a $\bpol{\Gs^+}$-morphism. We show that it satisfies~\eqref{eq:wgone}. By hypothesis, there exists a finite set \Hb of languages in \pol{\Gs^+} such that for every $s \in M$, $\alpha\inv(s)$ is a Boolean combination of languages in~\Hb. Since \pol{\Gs^+} is a \pvari, Proposition~\ref{prop:genocm} yields a \pol{\Gs^+}-morphism $\eta: A^* \to (N,\leq)$ recognizing all $H \in \Hb$. Moreover, Lemma~\ref{lem:pairsm} yields a \Gs-morphism $\beta: A^* \to G$ such that for every $u,v \in A^*$, if $\beta(u) = \beta(v)$, then $(\eta(u),\eta(v)) \in N^2$ is a \Gs-pair for $\eta$. We know that $G$ is a group by Lemma~\ref{lem:gmorph}. Finally, we let $n = \omega(M) \times \omega(N) \times \omega(G)$.

  We now prove that~\eqref{eq:gone} holds. Let $e,f \in E(S)$ and $q,r,s,t \in M$ such that $(q,s)$ is a \Gs-pair. We prove that $(eqfre)^{\omega}(esfte)^{\omega+1} = (eqfre)^{\omega}qft(esfte)^{\omega}$. We let $z'_e,z'_f \in A^+$ such that $\alpha(z'_e) = e$ and $\alpha(z'_f) = f$ (recall that $S = \alpha(A^+)$). Moreover, we let $z_e = (z'_e)^n$ and $z_f = (z'_f)^n$. Clearly, $\alpha(z_e) = e$ and $\alpha(z_f) = f$. Moreover, $\beta(z_e) = \beta(z_f) = 1_G$ by definition of $n$ since $G$ is a group.   Since $\beta: A^* \to G$ is a \Gs-morphism, Lemma~\ref{lem:pairsm} yields $u,x \in A^*$ and $g \in G$ such that $\beta(u) = \beta(x) = g$, $\alpha(u) = q$ and $\alpha(x) = s$. Moreover, let $v,y \in A^*$ such that $\alpha(v) = r$ and $\alpha(y) = t$. We let  $u' = z_euz_f$, $v' =  vz_e(z_euz_fvz_e)^{n-1}$, $x' = z_exz_f$ and \mbox{$y' = yz_e (z_exz_fyz_e)^{n-1}$}. We have $\beta(u') = \beta(u) = g$ and \mbox{$\beta(x') = \beta(x) = g$}. By definition of $n$, one may also verify that $\beta(u'v') = \beta(x'y') = 1_G$. Therefore, $\beta(v') = \beta(y') = g\inv$. It follows that $\beta(u'y') = 1_G$ and $\beta(v'x') = 1_G$. Hence, by definition of $\beta$, $(1_N,\eta(u'y'))$ and $(1_N,\eta(v'x'))$ are \Gs-pairs for $\eta$. Since $\eta$ is a \pol{\Gs^+}-morphism and $\eta(z_e),\eta(z_f) \in E(\eta(A^+))$ by definition of $n$, it then follows from Theorem~\ref{thm:polgp} that $\eta(z_e) \leq \eta(z_euy'z_e)$ and $\eta(z_f) \leq \eta(z_fv'xz_f)$. One may now multiply to obtain the following inequalities:
  \[
    \begin{array}{lll}
      \eta((u'v')^n (x'y')^{n+1}) & \leq &\eta((u'v')^n u'y'(x'y')^{n+1}). \\
      \eta((u'v')^n u'y' (x'y')^{n}) & \leq &\eta((u'v')^n u'v'x'y'(x'y')^{n}).
    \end{array}
  \]
  One may verify from the definitions of $u',v',x',y'$ and $n$ that when put together, these inequalities imply that $\eta((z_euz_fvz_e)^n (z_exz_fyz_e)^{n+1}) = \eta((z_euz_fvz_e)^n uz_fy(z_exz_fyz_e)^{n})$. Since $\eta$ recognizes all $H \in \Hb$, this yields $(z_euz_fvz_e)^n (z_exz_fyz_e)^{n+1} \in H \Leftrightarrow (z_euz_fvz_e)^n uz_fy(z_exz_fyz_e)^{n} \in H$ for every $H \in \Hb$. Finally, since all languages recognized by $\alpha$ are Boolean combination of languages in~\Hb, this implies that these two words have the same image under $\alpha$. By definition, this exactly says that $(eqfre)^{\omega}(esfte)^{\omega+1} = (eqfre)^{\omega}qft(esfte)^{\omega}$ as desired.

  \medskip

  We turn to the converse implication. Assume that $\alpha$ satisfies~\eqref{eq:wgone}. We prove that $\alpha$ is a \bpol{\Gs^+}-morphism. Lemma~\ref{lem:pairsm} yields a \Gs-morphism $\beta: A^* \to G$ such that for every $u,v \in A^*$, if $\beta(u) = \beta(v)$, then $(\alpha(u),\alpha(v))$ is a \Gs-pair. We write $L = \beta\inv(1_G) \in \Gs$. By hypothesis on \Gs, $L$ is a group language. Moreover, we have $\veps \in L$ by definition. Given a finite set of languages \Kb, and $s,t \in M$, we say that \Kb is \emph{$(s,t)$-safe} if for every $K \in \Kb$ and $w,w' \in K$, we have $s\alpha(w)t = s\alpha(w')t$. The argument is based on the following lemma.

  \begin{lemma}\label{lem:wbpgmain}
    Let $s,t \in M$. There exists an $(s,t)$-safe \bpol{\Gs^+}-cover of $L$.
  \end{lemma}

  We first apply Lemma~\ref{lem:wbpgmain} to prove that every language recognized by $\alpha$ belongs to \bpol{\Gs^+}. We apply it in the case when $s = t=1_M$. This yields a \bpol{\Gs^+}-cover $\Kb_L$ of $L$ which is $(1_M,1_M)$-safe. We use it to build a \bpol{\Gs}-cover \Kb of $A^*$ which is $(1_M,1_M)$-safe. Since $L \in \Gs$ and $\veps \in L$, Proposition~\ref{prop:pgcov} yields a cover \Pb of $A^*$ such that every $P \in \Pb$, there exist $n \in \nat$ and $a_1,\dots,a_n \in A$ such that $P = La_1L \cdots a_nL$. We cover each $P \in \Pb$ independently. Consider a language $P \in \Pb$. By definition, $P = La_1L \cdots a_nL$ for $a_1,\dots,a_n \in A$. Since $L \in \Gs$ and $\Kb_L$ is a \bpol{\Gs^+}-cover of $L$,  Proposition~\ref{prop:bconcat} yields a \bpol{\Gs^+}-cover $\Kb_P$ of $P =  La_1L \cdots a_nL$ such that for every $K \in \Kb_P$, there exist $K_0,\dots,K_n \in \Kb_L$ satisfying $K \subseteq K_0a_1K_1 \cdots a_nK_n$. Since $\Kb_L$ is $(1_M,1_M)$-safe, it is immediate that $\Kb_P$ is $(1_M,1_M)$-safe as well. Finally, since \Pb is a cover of $A^*$, it is now immediate that $\Kb=\bigcup_{P\in\Pb}\Kb_P$ is a $(1_M,1_M)$-safe \bpol{\Gs^+}-cover of $A^*$. Since \Kb is $(1_M,1_M)$-safe, we know that for every $K \in \Kb$, there exists $s\in M$ such that $K \subseteq \alpha\inv(s)$.  Hence, since \Kb is a cover of $A^*$, it is immediate that for every $F \subseteq M$, the language $\alpha\inv(F)$ is a union of languages in \Kb. By closure under union, it follows that $\alpha\inv(F) \in \bpol{\Gs^+}$. This exactly says that all languages recognized by $\alpha$ belong to \bpol{\Gs^+}.

  \smallskip

  It remains to prove Lemma~\ref{lem:wbpgmain}. We define a preorder on $M^2$ that we shall use as an induction parameter. Let $(s,t),(s',t') \in M^2$. We write $(s,t)\leqslant_{L}^+ (s',t')$ if either $(s,t) = (s',t')$ or there exist $x,y \in A^*$ and $e \in E(S)$ such that $xy \in L$, $\alpha(x)e = \alpha(x)$, $e\alpha(y) = \alpha(y)$, $s'=s\alpha(x)$ and $t' =\alpha(y)t$. It is immediate by definition that $\leqslant_{L}^+$ is reflexive. Let us verify that it is transitive. Let $(s,t),(s',t'),(s'',t'') \in M^2$ such that $(s,t) \leqslant_{L}^+ (s',t')$ and $(s',t') \leqslant_{L}^+ (s'',t'')$. We show that $(s,t) \leqslant_{L}^+ (s'',t'')$. If either $(s,t) = (s',t')$ or $(s',t') = (s'',t'')$, this is immediate. Otherwise, we have $x,y,x',y' \in A^*$ and $e,e' \in E(S)$ such that $xy,x'y' \in L$, $\alpha(x)e = \alpha(x)$, $e\alpha(y) = \alpha(y)$, $\alpha(x')e' = \alpha(x')$, $e'\alpha(y') = \alpha(y')$, $s'=s\alpha(x)$, $t' =\alpha(y)t$, $s''=s'\alpha(x')$ and $t'' =\alpha(y')t'$.  Since $L = \beta\inv(1_G)$, it is immediate that $xx'y'y \in L$. Moreover, $\alpha(xx')e' = \alpha(xx')$, $e'\alpha(y'y) = \alpha(y'y)$, $s''=s\alpha(xx')$ and $t'' =\alpha(y'y)t$. Hence, $(s,t) \leqslant_{L}^+ (s'',t'')$ as desired.

  We may now start the proof. Let $s,t \in M$. We construct a \bpol{\Gs^+}-cover \Kb of $L$ which is $(s,t)$-safe. We proceed by induction on the number of pairs $(s',t') \in M^2$ such that $(s,t) \leqslant_{L}^+ (s',t')$. The base case and the inductive step are handled simultaneously. First, we define a language $H \subseteq L$. Let $w \in L$. We say  \emph{$w$ stabilizes $(s,t)$} if $w = \veps$ or $w \in A^+$ and there exists $n \geq 1$, an $\alpha$-guarded decomposition $(w_1,\dots,w_{n+1})$ of $w$, an index $1\leq i\leq n$, \mbox{$x_1,\dots,x_i, y_i,\dots,y_n \in A^*$} and $e \in E(S)$ which satisfy the following conditions:
  \begin{itemize}
    \item $x_1,\dots,x_{i-1},x_iy_i, y_{i+1},\dots,y_n \in L$, and,
    \item $s\alpha(w_1x_1 \cdots w_ix_i)e = s$, and,
    \item $e\alpha(y_iw_{i+1} \cdots y_nw_{n+1})t = t$.
  \end{itemize}
  We let $H \subseteq L$ as the language of all words $w \in L$ which do \emph{not} stabilize $(s,t)$. Observe that by definition, we have $\veps \not\in H$. We first use induction to build an $(s,t)$-safe \bpol{\Gs^+}-cover of $H$. Then, we complete it to obtain the desired \bpol{\Gs^+}-cover of $L$. It may happen that $H$ is empty. In this case, we do not need induction: it suffices to use $\emptyset$ as this \bpol{\Gs^+}-cover.

  \smallskip

  We let $P \subseteq M^2$ as the set of all $(s',t') \in M^2$ such that $(s,t) \leqslant_{L}^+ (s',t')$ and \mbox{$(s',t') \not\leqslant_{L}^+ (s,t)$}. We define $\ell = |P|$ and write $P = \{(s'_1,t'_1), \dots, (s'_\ell,t'_\ell)\}$. For every $i \leq \ell$, we may apply induction in the proof of Lemma~\ref{lem:wbpgmain} to obtain a \bpol{\Gs^+}-cover $\Kb_i$ of $L$ which is $(s'_i,t'_i)$-safe. We define $\Kb_L = \left\{L \cap K_1 \cap \cdots \cap K_\ell \mid \text{$K_i \in \Kb_{i}$ for every $i \leq \ell$} \right\}$. Since $L \in \Gs$,  it is immediate by definition that $\Kb_L$ is a \bpol{\Gs^+}-cover of $L$ which is $(s',t')$-safe for every $(s',t') \in P$. We use it to build $\Kb_H$.

  \begin{lemma}\label{lem:wginduc}
    There exists an $(s,t)$-safe \bpol{\Gs^+}-cover $\Kb_H$ of $H$.
  \end{lemma}

  \begin{proof}
    Since $L$ is a group language such that $\veps \in L$ and $\veps \not\in H$, Proposition~\ref{prop:pgpcov} yields a cover \Ub of $H$ such that each $U \in \Ub$ is of the form $U = w_1L \cdots w_nLw_{n+1}$ where $(w_1,\dots,w_{n+1})$ is an $\alpha$-guarded decomposition of a word $w\in H$. For each $U \in \Ub$, we build an $(s,t)$-safe \bpol{\Gs^+}-cover $\Kb_U$ of~$U$. As \Ub is a cover of $H$, it will then suffice to define $\Kb_H$ as the union of all covers $\Kb_U$. We fix $U \in \Ub$.

    By definition of \Ub, $U = w_1L \cdots w_nLw_{n+1}$ where $(w_1,\dots,w_{n+1})$ is an $\alpha$-guarded decomposition of a word $w\in H$. Since $L\in\Gs$, $\veps \in L$ and $\Kb_L$ is a \bpol{\Gs^+}-cover of $L$ by hypothesis, Corollary~\ref{cor:bconcat} yields a \bpol{\Gs^+}-cover $\Kb_U$ of  $U$ such that for each $K \in \Kb_U$, we have $K \subseteq w_1K_1 \cdots w_n K_iw_{n+1}$ for $K_1,\dots,K_n \in \Kb_L$. It remains to show that $\Kb_U$ is $(s,t)$-safe. We fix $K \in \Kb_U$ as described above and $u,u' \in K$. We show that $s\alpha(u)t = s\alpha(u')t$.  If $n = 0$, then $K \subseteq \{w_1\}$. Hence $u = u' = w_1$ and the result is immediate. Assume now that $n \geq 1$. We get $u_i,u'_i \in K_i$ for $1 \leq i \leq n$ such that $u = w_1u_1 \cdots w_nu_nw_{n+1}$ and $u' = w_1u'_1 \cdots w_nu'_nw_{n+1}$. For $1 \leq i \leq n$, we write $x_i = w_1u_1w_2 \cdots u_{i-1}w_i$ and $x'_i = w_1u'_1w_2 \cdots u'_{i-1}w_i$ (when $i = 1$, $x_1 = x'_1 = w_1$). Moreover, we write $y_i = w_{i+1}u_{i+1} \cdots w_nu_nw_{n+1}$ and $y'_i = w_{i+1}u'_{i+1} \cdots w_nu'_nw_{n+1}$ ($y_n= y'_n = w_{n+1}$). Observe that for $1 \leq i \leq n$, we have $x_iu'_iy'_i = x_{i-1}u_{i-1}y'_{i-1}$. Hence, it suffices to prove that $s\alpha(x_iu_iy'_i)t = s\alpha(x_iu'_iy'_i)t$ for $1 \leq i \leq n$. It will then be immediate by transitivity that $s\alpha(x_nu_ny'_n)t = s\alpha(x_1u'_1y'_1)t$, \emph{i.e.} $s\alpha(u)t = s\alpha(u')t$ as desired. We fix $i$ such that $1 \leq i\leq n$ for the proof.

    We have $u_i,u'_i \in K_i$ by hypothesis. Hence, since $K_i \in \Kb_L$ which is $(s',t')$-safe for every $(s',t')\in P$, it suffices to prove that $(s\alpha(x_i),\alpha(y'_i)t) \in P$. There are two conditions to verify. First, we show that $(s,t) \leqslant_{L}^+ (s\alpha(x_i),\alpha(y'_i)t)$. By definition of $\leqslant_{L}^+$, this boils down to proving that $x_iy'_i \in L$ and there exists $e \in E(S)$ such that $\alpha(x_i)e = \alpha(x_i)$ and $e\alpha(y'_i) =\alpha(y'_i)$. We get $e \in E(S)$ by definition of $x_i,y'_i$ as $(w_1,\dots,w_{n+1})$ is an $\alpha$-guarded decomposition. Moreover, $u_j,u'_j \in K_j$ for every $j \leq n$ and since $K_j \in \Kb_L$, it follows that $u_j,u'_j \in L$ for every $j \leq n$ by definition of $\Kb_P$. Hence, $\beta(u_j) = \beta(u'_j) = 1_G$ since $L = \beta\inv(1_G)$. By definition of $x_i$ and $y'_i$, we obtain $\beta(x_i) = \beta(w_1 \cdots w_i)$ and $\beta(y'_i) = \beta(w_{i+1} \cdots w_{n+1})$. This yields $\beta(x_iy'_i) = \beta(w_1 \cdots w_{n+1})$. Finally, since $(w_1,\dots,w_{n+1})$ is an $\alpha$-guarded decomposition of $w \in H$, we have $w_1 \cdots w_{n+1} = w \in H \subseteq L$. Since $L$ is recognized by $\beta$, it follows that $x_iy'_i \in L$ as desired.

    We now prove that $(s\alpha(x_i),\alpha(y'_i)t) \not\leqslant_{L}^+ (s,t)$. We proceed by contradiction. Assume that $(s\alpha(x_i),\alpha(y'_i)t) \leqslant_{L}^+ (s,t)$. We prove that $w$ stabilizes $(s,t)$, contradicting the hypothesis that $w \in H$. For this purpose, we exhibit $x,y \in A^*$ and $f \in E(S)$ such that $xy \in L$, $(s,t) = (s\alpha(x_ix),\alpha(yy'_i)t)$, $\alpha(x_ix)f = \alpha(x_ix)$ and $f\alpha(yy'_i) = \alpha(yy'_i)$. Since $u_j,u'_j \in L$ for every $j \leq n$, this clearly implies that $w$ stabilizes $(s,t)$ by definition of $x_i$ and $y'_i$ from that $\alpha$-guarded decomposition $(w_1,\dots,w_{n+1})$ of $w$. It remains to exhibit the appropriate $x,y \in A^*$ and $f \in E(S)$. Since $(s\alpha(x_i),\alpha(y'_i)t) \leqslant_{L}^+ (s,t)$, there are two cases. When $(s\alpha(x_i),\alpha(y'_i)t) = (s,t)$, it suffices to choose $x = y = \veps$ and $f = e$. Otherwise, there exist  $x,y \in A^*$ and $f \in E(S)$ such that $(s,t) = (s\alpha(x_ix),\alpha(yy'_i)t)$, $\alpha(x)f = \alpha(x)$ and $f\alpha(y) = \alpha(y)$. Hence, the result is also immediate.
  \end{proof}

  We now define the desired $(s,t)$-safe \bpol{\Gs^+}-cover \Kb of $L$. Consider the \bpol{\Gs^+}-cover $\Kb_H$ of $H$ provided by Lemma~\ref{lem:wginduc}. We let $K_\bot = L \setminus (\bigcup_{K \in \Kb_H} K)$. Finally, we let $\Kb = \{K_\bot\} \cup \Kb_H$. It is immediate that \Kb is a \bpol{\Gs^+}-cover of $L$ since \bpol{\Gs^+} is a Boolean algebra (recall that $L \in \Gs$). It remains to verify that \Kb is $(s,t)$-safe. Since we already know that $\Kb_H$ is $(s,t)$-safe, it suffices to prove that for every $w,w' \in K_{\bot}$, we have $s\alpha(w)t = s\alpha(w')t$. We actually show that $s\alpha(w)t = st$ for every $w \in K_\bot$. Since this is immediate when $w = \veps$, we assume that $w \in A^+$.

  By definition of $K_\bot$, we have $w \not\in K'$ for every $K' \in \Kb_H$. Since $\Kb_H$ is a cover of $H$, it follows that $w \not\in H$ which means that $w$ stabilizes $(s,t)$ by definition of $H$. Since $w \neq \veps$, we get an $\alpha$-guarded decomposition $(w_1,\dots,w_{n+1})$ of $w$, some index $1\leq i\leq n$, $x_1,\dots,x_i, y_i,\dots,y_1 \in A^*$ and $e \in E(S)$ such that we have $x_1,\dots,x_{i-1},x_iy_i, y_{i-1},\dots,y_n \in L$, $s\alpha(w_1x_1\cdots w_ix_i)e = s$ and $e\alpha(y_iw_{i+1} \cdots y_nw_n)t = t$. We let $u = w_1 \cdots w_i$ and $v = w_{i+1}\cdots w_{n+1}$. We show that $s = s \alpha(ux_i)e$ and $t = e\alpha(y_iv)t$ (note that since $e$ is an idempotent, this also implies $s = se$ and $t = et$) Let us first assume that this holds and explain why this implies $st = s\alpha(w)t$.

  Since $(w_1,\dots,w_{n+1})$ is an $\alpha$-guarded decomposition, there exist an idempotent $f \in E(S)$ such that $\alpha(w_i)f=\alpha(w_i)$ and $f\alpha(w_{i+1})= \alpha(w_{i+1})$. By definition of $u$ and $v$, we have $\alpha(u)f = \alpha(u)$ and $f\alpha(v) = \alpha(v)$. Clearly, we have $uv = w$. Thus, since $w \in L = \beta\inv(1_G)$, we have $\beta(u)\beta(v) = 1_G$. Let $p = \omega(G)$. We have $1_G = \beta((y_iv)^p)$. Thus, since $G$ is a group, it follows that $\beta(u) = \beta((y_iv)^{p-1}y_i)$. By definition of $\beta$, it follows that $(\alpha(u),\alpha((y_nv)^{p-1}y_i))$ is a \Gs-pair. Let $q = \alpha(u)$, $r = \alpha(x_i)$, $q' = \alpha((y_nv)^{p-1}y_i)$ and $r' = \alpha(v)$. Since we just proved that $(q,q')$ is a \Gs-pair, we obtain from~\eqref{eq:wgone} that,
  \begin{equation} \label{eq:mbp:wproof}
    (eqfre)^\omega (eq'fr'e)^{\omega+1} =   (eqfre)^\omega qfr' (eq'fr'e)^{\omega}.
  \end{equation}
  Since $\alpha(u)f = \alpha(u)$, we have $eqfre = e\alpha(ux_i)e$ and $qfr' = \alpha(uv) = \alpha(w)$. Moreover, since $f\alpha(v) = \alpha(v)$, we have $eq'fr'e = e\alpha((y_iv)^p)e$. Hence, since we have $s = s \alpha(ux_i)e = se$ and $t = e\alpha(y_iv)t  = et$, it is immediate that $seqfre = s$ and $eq'fr'et = t$. We may now multiply by $s$ on the left and $t$ on the right in~\eqref{eq:mbp:wproof} to obtain $st = sqfr't = s \alpha(w)t$ as desired.

  It remains to prove that $s = s \alpha(ux_i)e$ and $t = e\alpha(y_iv)t$. We concentrate on $s = s \alpha(ux_i)e$ (the other equality is symmetrical and left to the reader). For every $j$ such that $1 \leq j \leq i$, we write $r_j = \alpha(w_jx_j \cdots w_ix_i)e$ and $u_j = w_{1} \cdots w_{j-1}$ (we let $u_1 = \veps$). We use induction on $j$ to prove that $s = s \alpha(u_j)r_j$ for $1 \leq j \leq i$. This concludes the argument: when $j = i$, we get $s = s \alpha(w_1 \cdots w_{i-1}w_ix_i)e$. Since $u = w_1 \cdots w_i$, this exactly says that $s = s \alpha(ux_i)e$ as desired. The case $j = 1$ is immediate by definition: we have $s\alpha(w_1x_1\cdots w_ix_i)e = s$. Thus, we now assume that $2 \leq j \leq i$. Since $(w_1,\dots,w_{n+1})$ is an $\alpha$-guarded decomposition, there exist an idempotent $f \in E(S)$ such that $\alpha(w_{j-1})f=\alpha(w_{j-1})$ and $f\alpha(w_{j})= \alpha(w_{j})$. By definition of $u_j$ and $r_j$, we have $\alpha(u_j)f = \alpha(u_j)$ and $fr_j = r_j$. Moreover, since $x_{j-1} \in L$ and $L = \beta\inv(1_G)$, we have $\beta(x_{j-1}) = \beta(\veps) = 1_G$. By definition of $\beta$, it follows that $(\alpha(x_{j-1}),1_M)$ is a \Gs-pair. Hence, we may apply~\eqref{eq:wgone} for $q = \alpha(x_{j-1})$, $r = r_j\alpha(u_j)$ and $s = t = 1_M$ to obtain,
  \begin{equation} \label{eq:mbp:wgrpeq}
    (f\alpha(x_{j-1})fr_j\alpha(u_{j})f)^\omega =  (f\alpha(x_{j-1})fr_j\alpha(u_{j})f)^\omega \alpha(x_{j-1})f.
  \end{equation}
  It follows from induction that $s = s \alpha(u_{j-1})r_{j-1}$. Moreover, it is immediate from the definitions that we have  $\alpha(u_{j-1})r_{j-1} = \alpha(u_j)\alpha(x_{j-1})r_{j} = \alpha(u_j)f\alpha(x_{j-1})fr_{j}$. Therefore, we get,
  \[
    \begin{array}{llll}
      s & = & s  \alpha(u_j)f\alpha(x_{j-1})fr_{j} &\\
        & = & s  (\alpha(u_j)f\alpha(x_{j-1})fr_{j})^{\omega+1}        &                                                \\
        & = & s \alpha(u_{j}) (f\alpha(x_{j-1})fr_j\alpha(u_{j})f)^\omega \alpha(x_{j-1})fr_j&  \\
        & = & s \alpha(u_{j}) (f\alpha(x_{j-1})fr_j\alpha(u_{j})f)^\omega r_j&  \quad \text{by~\eqref{eq:mbp:wgrpeq}}\\
        & = & s (\alpha(u_{j})f\alpha(x_{j-1})fr_{j})^{\omega} \alpha(u_{j})fr_j& \\
        & = & s \alpha(u_{j})fr_j. &
    \end{array}
  \]
  This exactly says that $q = s \alpha(u_{j})r_j$ which completes the proof.
\end{proof}

\section{Conclusion}
\label{sec:conc}
We presented generic algebraic characterizations for classes of the form \bpol{\Gs} and \bpol{\Gs^+} when \Gs is a \vari of group languages. They imply that membership is decidable for these two classes as soon as separation is decidable for the input class \Gs. The most natural follow-up question is whether these two characterizations can be generalized to encompass all classes \bpol{\Cs} where \Cs is an \emph{arbitrary} \vari and obtain a characterization similar to the one provided by Theorem~\ref{thm:polc} for \pol{\Cs}. This is a difficult question. In particular, it seems unlikely that \bpol{\Cs}-membership boils down to \Cs-separation in the general case. Indeed,  a specialized characterization for the class \bpol{\bpol{\stzer}} is known~\cite{pzjacm19}. Yet, deciding it involves looking at more general question than \bpol{\stzer}-separation.

\bibliographystyle{abbrv}

\end{document}